\documentclass[sort&compress
    ,draft           
    ,numberedheadings 
    ,cmfonts          
    ]
  {aipproc}

\usepackage[draft=false,breaklinks=true, debug=true, colorlinks=true, linkcolor=blue, citecolor=red, urlcolor=blue]{hyperref}

\usepackage{subfigure}
\usepackage{amsmath}
\usepackage{amssymb}

\date{J.P. Lansberg \etal}

\newcommand{\be}{\begin{equation}}
\newcommand{\ee}{\end{equation}}
\newcommand{\bea}{\begin{eqnarray}}
\newcommand{\eea}{\end{eqnarray}}

\newcommand{\jpsi}{{\ensuremath{J/\psi}}}
\newcommand{\ups}{{\ensuremath{\Upsilon}}}

\newcommand{\ptjpsi}{{\ensuremath{\rm P_{T}^{J/\psi}}}}

\newcommand{\ccbar}{\ensuremath{{\rm c\bar{c}}}}

\providecommand{\elel}{e^+e^-}
\providecommand{\mumu}{\mu^+\mu^-}

\newcommand{\eqs}[1]{\begin{equation} \begin{split} #1\end{split} \end{equation} }

\newcommand{\eg}{{e.g.}}
\newcommand{\etal}{{\it et al.}}

\newcommand{\cf}[1]{{Fig.~\ref{#1}}}

\newcommand{\sqrts}{\sqrt{s}}
\newcommand{\sqrtsnn}{\sqrt{s_{_{NN}}}}

\providecommand{\jpsi}{J/\psi}

\providecommand{\ups}{\Upsilon}

\providecommand{\qqbar}{Q\bar{Q}}
\providecommand{\ccbar}{c\bar{c}}

\providecommand{\str}{{\sc starlight }}

\providecommand{\gaga}{\gamma\,\gamma}

\providecommand{\gA}{\gamma\,A}

\providecommand{\pp}{{\it pp}}
\providecommand{\pA}{{\it pA}}
\providecommand{\pPb}{{\it pPb}}

\providecommand{\dAu}{{\it dAu}}
\providecommand{\AaAa}{{\it AA}}
\providecommand{\AuAu}{{\it AuAu}}
\providecommand{\PbPb}{{\it PbPb}}

\layoutstyle{6x9}

\begin{document}

\title{Perspectives on heavy-quarkonium production at the LHC}

\classification{14.40.Gx,13.85.Ni,25.75.-q,25.20.Lj,12.38.Mh}
\keywords      {Quarkonium production, LHC}
\author{J.P. Lansberg}{
  address={Institut f\"ur Theoretische Physik, Universit\"at Heidelberg, Philosophenweg 19, \\ D-69120 Heidelberg, Germany}
}
\author{A. Rakotozafindrabe}{
  address={IRFU/SPhN, CEA Saclay, 91191 Gif-sur-Yvette Cedex, France}
}
\author{P. Artoisenet}{
  address={Center for Particle Physics and Phenomenology (CP3), Universit\'e catholique de Louvain,
B-1348 Louvain-la-Neuve, Belgium}
}
\author{D. Blaschke}{
  address={Instytut Fizyki Teoretycznej, Uniwersytet Wroc\l awski, 50-204 Wroc\l aw, Poland}
}
\author{J. Cugnon}{
  address={D\'epartement d'Astrophysique, de G\'eophysique et d'Oc\'eanographie, Universit\'e de  Li\`ege, 
all\'ee du 6 Ao\^ut 17, b\^at. B5
, B-4000 Li\`ege, Belgium}
}
\author{D. d'Enterria}{
  address={CERN, PH-EP
CH-1211 Geneva 23, Switzerland}
}
\author{A. C. Kraan}{
  address={Istituto Nazionale di Fisica Nucleare di Pisa, Largo Pontecorvo 3, 56100, Pisa, Italy}
}
\author{F. Maltoni}{
  address={Center for Particle Physics and Phenomenology (CP3), Universit\'e catholique de Louvain,
B-1348 Louvain-la-Neuve, Belgium}
}
\author{D. Prorok}{
  address={Instytut Fizyki Teoretycznej, Uniwersytet Wroc\l awski, 50-204 Wroc\l aw, Poland}
}
\author{H. Satz}{
  address={Fakult\"at f\"ur Physik, Universit\"at Bielefeld, Universit\"atsstra\ss e 25, D-33615n Bielefeld, Germany}
}

\begin{abstract}

We summarise the perspectives on heavy-quarkonium production at the LHC, both for proton-proton and heavy-ion runs, 
as emanating from the round table held at the HLPW
2008 Conference. The main topics are: present experimental and theoretical
knowledge,  experimental capabilities, open questions, recent theoretical advances
and potentialities linked to some new observables.

\end{abstract}

\maketitle

\renewcommand{\thefootnote}{\fnsymbol{footnote}}
\footnotetext{E-mails:\\ lansberg@tphys.uni-heidelberg.de, andry.rakotozafindrabe@cea.fr, blaschke@ift.uni.wroc.pl,\\ pierre.artoisenet@uclouvain.be, J.Cugnon@ulg.ac.be, dde@cern.ch, Aafke.Kraan@cern.ch,\\ fabio.maltoni@uclouvain.be,prorok@ift.uni.wroc.pl,satz@physik.uni-bielefeld.de}
\renewcommand{\thefootnote}{\arabic{footnote}}

\tableofcontents


\section{Introduction}

With the start-up of the LHC approaching, it is certainly expedient to
make an overview on what we currently know on quarkonium production,
both in proton-proton and heavy-ion collisions and on what we can
expect from analyses to be carried at the LHC. Heavy-quarkonium
production mechanism has always been -- and still is -- a subject of
debate (for reviews
see~\cite{Brambilla,Lansberg:2006dh,Bedjidian:2004gd,Vogt:1999cu,Kopeliovich:2003cn}). Heavy
quarkonia have been often suggested as ideal probes in studies and
analyses of complex phenomena. However, reality was later found to be much less
simple than initially thought.  A well known example is the suggestion
to measure the suppression of $J/\psi$ production in heavy-ion
collision as a smoking-gun signature of the creation of the quark
gluon plasma (QGP)~\cite{Matsui:1986dk}. However, cold nuclear matter
effects, such as shadowing, energy loss, absorption, etc. , were shown
to play an important role and had to be considered in the interpretation of 
the experimental measurements. Furthermore, effects of the
successive dissociations of higher-excited states which can decay into
$J/\psi$ ($\psi'$, $\chi_c$) had to be taken into account, as well as
more specific issues related to the description of the plasma itself.

In fact, even in a much ``cleaner''environment, such as in \pp\
collisions, understanding quarkonium production has been a challenge
since the first measurements by the CDF Collaboration of the {\it
direct} production of $J/\psi$ and $\psi'$ at $\sqrt{s}=1.8$
TeV~\cite{Abe:1997jz,Abe:1997yz}.  It is fair to say that, at present,
a consistent theoretical picture that predicts both cross sections and
the polarisation measurements for charmonium at the
Tevatron~\cite{Abulencia:2007us}, along with the cross section from
PHENIX at RHIC~\cite{Adare:2006kf} is not available.  For instance,
the long-standing prediction of Non Relativistic
QCD~\cite{Bodwin:1994jh} (NRQCD) on the transverse polarisation of
$\psi$'s at high transverse momentum is not supported by the data.  The
most natural interpretation of such flagrant failure of NRQCD is that
the charmonium system is too light for relativistic effects to be
neglected.

Indications that this might be indeed the case come from the agreement
between theory and the available experimental data for $\Upsilon$
production in \pp\ (and inclusive decays). In this case, relativistic 
corrections are expected to be less important  and the leading state
in the Fock expansion, i.e.  the heavy-quark pair in a colour singlet
$^3S_1$ to be dominant. The latest NLO predictions~\cite{NNLO}
which include some of the important NNLO $\alpha_S^5$ corrections, show a
satisfactory agreement with the data coming from the Tevatron
~\cite{Acosta:2001gv,Abazov:2005yc}. Once again, much is expected from
polarisation measurements at the Tevatron and the LHC to confirm that at
least bottomonium predictions are well under control.

In \pA\ collisions, LHC measurements will certainly be of greatest
importance to pinpoint the size of shadowing effects in the small $x$
region which can also be studied via electromagnetic (aka. ``ultraperipheral'')
\AaAa\ collisions. Furthermore, data will allow to understand the
absorption mechanisms at high-energy and subsequently to gain insights
of the different formation time of the various heavy quarkonia.

Finally, nucleus-nucleus collisions at the LHC will be the
long-awaited ideal laboratory for the study of the QGP. Unprecedented
temperatures will be reached; in conjunction with the high-luminosity
beams, bottomonia will be at last promoted as a practical probe for
the QGP formation. With the good news from higher-order QCD-correction
studies centered on $\Upsilon$, we are certainly at the beginning of
very exciting discovery years.

\section{Experimental capabilities}

\subsection{ALICE \protect\small (by Andry M. Rakotozafindrabe)}
\label{sec:alice}

ALICE is the LHC dedicated heavy-ion experiment. Its main physics goal
is to study the properties of the hot and dense deconfined hadronic
matter which is expected to be created during the relativistic heavy
ion collisions. ALICE's primary interest into the production rate of
the heavy-quarkonium states lies into the fact that it can be used as a
sensitive probe to the formation of the quark-gluon plasma. 
At temperatures above the quarkonium
binding energy, the latter is foreseen to melt through colour
screening, inducing a suppression of the production rates. The dissociation
temperature $T_d$ pattern would be:

\begin{equation}
T_d[\psi'] \approx T_d[\chi_c] < T_d[ \Upsilon(3S) ] 
< T_d[ J/\psi ] \approx T_d[ \Upsilon(2S) ] < T_d[ \Upsilon(1S)) ]
\end{equation}

 which shows how quarkonium suppression can be used to estimate the
 temperature of the created QGP. Lower energy accelerators/colliders
 (SPS and RHIC) have indeed observed $J/\psi$ suppression. For the $\Upsilon$
 family, this will be only feasible at the LHC where the $b\bar{b}$
 production cross-section is quite sizeable, and where $T_d[
 \Upsilon(1S)]$ can be reached. Moreover, at nominal luminosity, central
 \PbPb\ collisions at the LHC ($\sqrtsnn$~=~5.5~TeV) are expected
 to produce about a hundred of $c\bar{c}$ pairs, which substantially
 increases the regeneration probability of secondary $J/\psi$'s. This
 emphasises the interest in measuring the $\Upsilon(2S)$
 production rate, since it is expected to benefit from a small
 regeneration probability (about only five $b\bar{b}$ pairs are
 expected in central \PbPb\ collisions). An important remark is that
 in-medium effects on charmonia can be studied once the feed-down from
 $B$ decays is properly subtracted ($B\rightarrow J/\psi + X$, expected
 to account for about $20\%$ of the total $J/\psi$ yield if no cold nuclear matter
 effect is considered). Obviously, results obtained in \AaAa\
 collisions must be benchmarked against the ones from \pp\
 collisions, but also \pA\ collisions in order to get the baseline
 for the cold nuclear effects.\\

 The LHC program plans \pp\
 running at $\sqrt{s}=14\mathrm{~TeV}$ for 8 months per year
 ($10^7 \mathrm{~s}$ effective time) followed by ion running for one
 month per year ($10^6 \mathrm{~s}$). ALICE can participate to \pp\
 running, but at a maximum allowed luminosity of
 $5\,10^{30}\mathrm{~cm}^{-2}\mathrm{~s}^{-1}$. The first
 heavy-ion run will be \PbPb\ collisions at
 $\sqrtsnn\,=\,5.5\mathrm{~TeV}$ at a luminosity of
 $5\,10^{25}\mathrm{~cm}^{-2}\mathrm{~s}^{-1}$, corresponding to
 $1/20^{\mathrm{th}}$ of the design luminosity. 
  One or two years of light-ion collisions ({\it ArAr}, with a luminosity up to
 $10^{30}\mathrm{~cm}^{-2}\mathrm{~s}^{-1}$), and one year
 of \pPb\ collisions at $\sqrtsnn=8.8\mathrm{~TeV}$ are also planned.

ALICE~\cite{Carminati:2004fp,Alessandro:2006yt} can detect quarkonia in
the dielectron channel at central rapidity ($|\eta|<0.9$), and in the
dimuon channel at forward rapidity ($-4 <\eta<-2.5$). In these
rapidity ranges, the corresponding probed Bjorken-$x$ approximately
goes from $10^{-2}$ to values as low as $10^{-5}$, 
providing additional insight into the PDFs and their
modification in the nuclear environment. The central barrel and the
muon arm are equipped with dedicated triggers on the individual
electron and muon transverse momentum. In order to select a lepton
pair from quarkonium decays, a low-$P_T$ trigger cut is applied on
individual leptons to remove most background sources:
$P_T>$3~GeV/$c$ for single electrons and $P_T>1 \,
(2)$~GeV/$c$ for single muons from charmonia (bottomonia). As a
consequence, it prevents the detection of charmonia with
$P_T<$~5~GeV/$c$ in the dielectron channel, whereas the
charmonia can be detected down to very low $P_T$ (about a hundred
of MeV/$c$) in the dimuon channel. The high $P_T$ reach is expected to be
$10 \, (20)$~GeV/$c$ for the $J/\psi$ into dielectrons
(dimuons) for a \PbPb\ run of one month at nominal luminosity. In
both channels, the expected mass resolution of about
$90\mathrm{~MeV}/c^2$ will be sufficient to resolve all the $\Upsilon$
states. The expected mass resolution for the $J/\psi$ is about
$30\,(70)\mathrm{~MeV}/c^2$ in the dielectron (dimuon) channel. The
central barrel has excellent secondary vertexing capabilities,
combined with particle identification. Therefore, prompt and
secondary $J/\psi$, from $B$ decays, can be distinguished at
central rapidities via a displaced vertex measurement. At
forward rapidity, this technique is not usable: the prompt $J/\psi$
yield has to be determined indirectly, by subtracting from the
measured yield the one expected from $B$ decay. The latter is inferred
from the single-muon $P_T$-spectra measurement with the cut
$P_T^{\mu}\gtrsim$~1.5~GeV/$c$ applied to all reconstructed
muons to maximise the beauty-signal significance. A fit technique is
then applied to extract the $P_T$ distribution of the muons from $B$
decays~\cite{Alessandro:2006yt}. Last, but not least, ALICE will be
able to measure the $J/\psi$ polarisation, both in \pp\ and \PbPb\
collisions.

\subsection{ATLAS and CMS \protect \small (by Aafke C. Kraan and David d'Enterria) }

The ATLAS~\cite{atlastdr} and CMS~\cite{cmstdr} experiments at the LHC are general 
purpose detectors designed to explore the physics at the TeV energy scale. The primary goals 
of the experiments are to reveal the electroweak symmetry breaking mechanism and 
provide evidence of physics beyond the Standard Model (SM) in proton-proton collisions at 
$\sqrts$~=~14 TeV. The two experiments are obviously extremely well suited to carry out 
Quantum-Chromodynamics studies in both \pp\ and \PbPb\ collisions.\\

The total cross section for \jpsi~and \ups~production at \pp\
collisions at 14 TeV are expected to be around 0.4 mb and 7 $\mu$b,
respectively, and for the higher mass states an order of magnitude
lower.  The $\jpsi$ and $\ups$ measurements in ATLAS and CMS focuse
normally in the dimuon decay channel ($\jpsi, \ups\rightarrow\mumu$)
with branching ratios of $\jpsi\rightarrow\mumu$ and
$\ups\rightarrow\mumu$ are 5.98\% and 2.48\% respectively.  On the one
hand, the high centre-of-mass energies (up to 14 TeV) and luminosities
(up to $10^{34}$ cm$^{-2}$s$^{-1}$) as well as the large-acceptance
muon spectrometers ($|\eta|<$2.5, full azimuth) available in the ATLAS
and CMS experiments will allow, unlike the ALICE and LHCb cases, for
quarkonium measurements at the LHC up to very large transverse
momenta.  On the other hand, the strong magnetic field and/or the large
material budget results in lower reconstruction efficiencies for small
transverse momentum muons (usually below $P_{T}\approx$~3~GeV/$c$).
Thus, the sensitivity for quarkonium measurements at low momentum is
inferior than that of the Tevatron or ALICE and LHCb experiments.
Only startup conditions, when the luminosities and muon trigger
thresholds will be low, will allow for some lower transverse momentum
measurements. 

Besides physics measurements, during detector
commissioning and startup, dimuon invariant masses of quarkonium
states will provide extremely useful candles for detector calibration
and alignment. Several physics analyses are planned at both ATLAS and CMS including:
\begin{itemize}
\item Differential inclusive cross section for \jpsi, \ups,
      $\psi(2S)$, $\ups(2S)$, $\ups(3S)$,
as well as the $\chi_c^0$, $\chi_c^1$, $\chi_c^2$ states in \pp, \pA\
and \AaAa\ collisions.
\item Polarisation measurements of these states.
\item More exclusive measurements aimed at understanding the
      underlying $Q\overline{Q}$ production mechanisms by looking, e.g., 
      at the associated hadronic activity.
\end{itemize}
In Table~\ref{tab:aafke} some relevant parameters for \jpsi's and
\ups's are given for the ATLAS and CMS
experiments~\cite{atlastdr,cmstdr,atlastalk, cmstalk}. For the
heavy-ion running, the performances of ATLAS~\cite{lebedev}
and CMS~\cite{cmshitdr} are very similar to the
proton-proton ones.

\begin{table}[ht!]
\begin{tabular}{||l|l|l||}
\hline
 & ATLAS & CMS\\\hline
$Q\overline{Q}$ trigger threshold & $P_T^{\mu_1}>6$~GeV/$c$, $P_T^{\mu_2}>4$~GeV/$c$, & $P_T^{\mu_1}>3$~GeV/$c$,
$P_T^{\mu_2}>3$~GeV/$c$\\ 
\jpsi~ mass resolution & $\sim$50 MeV/$c^2$  & $\sim$30 MeV/$c^2$ \\
$N_{\mathrm{events}}^{\mathrm{reco}}$ in 10 pb$^{-1}$ &  $2\times 10^5$ & $3 \times 10^5$\\
\ups~mass resolution & $\sim$170 MeV/$c^2$ &  $\sim$95 MeV/$c^2$ \\
$N_{\mathrm{events}}^{\mathrm{reco}}$ in 10 pb$^{-1}$ &  $5\times 10^4$ & $1\times 10^5$\\
\hline
\end{tabular}
\caption{Basic \jpsi~and \ups~reconstruction performances for ATLAS and CMS in \pp\ collisions at 14 TeV.\label{tab:aafke}}
\end{table}

\subsection{LHCb \protect\small (by David d'Enterria)}

The LHCb experiment~\cite{lhcb} at the LHC is mainly focused on the search of possible signals of new physics in
CP-violation and rare decays processes in the heavy-quark sector of the Standard Model (SM).
For this purpose, the experiment has been equipped with arguably the best capabilities for the detection
of $b$ and $c$ quarks produced at {\it forward rapidities} in proton-proton collisions at the LHC. 
Indeed, the LHCb detector -- with a single-arm configuration -- has excellent and varied particle detection 
and identification capabilities in the forward hemisphere. In particular, muons and electrons can be well 
measured in the pseudo-rapidity range $1.8<\eta<~4.9$. For comparison, in the same $\eta$ 
range ATLAS and CMS can only reconstruct electrons (in a reduced 2 $\lesssim\eta\lesssim$ 3 range), 
whereas ALICE can only measure muons (in a slightly more reduced acceptance: 2.5 $<\eta<$ 4, and 
only for proton-proton luminosities, 10$^{30}$ cm$^{-2}$s$^{-1}$, 100 times lower than those 
available for LHCb). Those characteristics make of LHCb an excellent apparatus to measure forward 
quarkonium production cross-sections and polarisation via $\jpsi,\ups\rightarrow \elel,\, \mumu$, 
including the excited states ($\psi(2S)$, $\Upsilon(\mathrm{2S})$, $\Upsilon(3S)$).

\section{Open issues}

\subsection{\pp\ collisions  \protect\small (by Andry M. Rakotozafindrabe and J.P. Lansberg)}

The underlying theory for {\it direct} and {\it prompt} $\psi$
is still under intense debate~\cite{Lansberg:2006dh,Brambilla}. 
Via the colour octet (CO) mechanism, NRQCD factorisation~\cite{Bodwin:1994jh} 
has been successful to explain some features of the charmonium hadroproduction. As illustrated by
the comparisons to CDF measurements in $p\bar p$~\cite{Abe:1997jz,Abe:1997yz} 
or to PHENIX old measurements in \pp~\cite{Adler:2003qs,Cooper:2004qe}), for 
$P_T\gtrsim 5$~GeV/$c$, it provides a good description of the $P_T$-differential 
cross-section for the direct~$J/\psi$ and $\psi'$, the cross-section being 
dominated by the gluon fragmentation into a colour-octet $^3S_1$ state. 
The latter mechanism leads to transversally polarised $J/\psi$ and $\psi'$. 

However, this is not seen by the CDF
experiment~\cite{Abulencia:2007us} which measured a slight
longitudinal polarisation for both the prompt~$J/\psi$ and
direct~$\psi'$ yield.  It is worth noting here that the feed-down from
$\chi_c$ can influence significantly the polarisation of the prompt
$J/\psi$ yield -- this was taken into account in the NRQCD-based
predictions~\cite{COM-pol}.
Moreover, the recent preliminary result
from PHENIX~\cite{ErmiasHP08} indicates a polarisation compatible with
zero for the total $J/\psi$ production at forward rapidity ($1.2 < |y|
< 2.2$), but with large uncertainties.

It is therefore not surprising to observe a renewed interest
in improving the present predictions for the colour singlet (CS) contribution, 
by computing the higher-order QCD (see section~\ref{sec:QCD-corrections}) 
corrections in $\alpha_S$,  or by ``softening'' some of the basic assumptions 
of the common approaches, as done in~\cite{schannelcut} via the 
consideration of the $s$-channel cut contribution.  On the one hand,
NLO~\cite{Campbell:2007ws,Artoisenet:2007xi} and part
of NNLO corrections~\cite{NNLO} significantly enhance the 
quarkonium yields\footnote{This sounds like a confirmation of the
study~\cite{Khoze:2004eu} which dealt in a simplified way with NNLO
corrections assimilable to LO BFKL contributions. However, this study
could not provide a prediction for the $P_T$ dependence: it only
predicted an enhancement of some NNLO corrections for large $s$.}.  In the
$\Upsilon(1S)$ and $\Upsilon(3S)$ cases, the
dominant NNLO corrections to the colour singlet~\cite{NNLO} suffice to
successfully describe the measured $P_T$-differential cross-section of
the direct yield~\cite{Acosta:2001gv,Abazov:2005yc}.  The polarisation
predictions for the latter cases seem quite encouraging considering
CDF~\cite{Acosta:2001gv} and $D\emptyset$~\cite{D0:2008za}
measurements.  Those corrections could be though still unable to bring
agreement with the measured $P_T$-differential cross-section of the
direct~$J/\psi$, but dedicated further studies are needed.

On the other hand, by including $s$-channel cut 
contributions~\cite{schannelcut} to the usual production CS production mechanism
one can reproduce the  $P_T$-spectra up to intermediate values of the $J/\psi$'s $P_T$, both at the
Tevatron and at RHIC, and  provide mostly longitudinally polarised $J/\psi$ at the Tevatron. 
However, as expected~\cite{schannelcut}, this approach underestimates the cross-section at 
large values of $P_T$. 

In summary, the theoretical status sounds clearer for the bottomonia
than for the charmonium production processes. Additional tests are
undoubtedly needed beyond the {\it mere} measurements of inclusive
cross section and polarisation at the LHC. For instance, the
hadroproduction of $J/\psi$ or $\Upsilon$ with a heavy-quark 
pair~\cite{Artoisenet:2007xi,Artoisenet:2008tc} appears to be a
new valuable tool to separately probe the CS contribution, at least
dominant at low-$P_T$ (below 15~GeV$/c$), as well as the study of the
hadronic activity around the quarkonium (see
section~\ref{sec:new-obs}).

\subsection{\pA\ collisions  \protect\small (by Andry M. Rakotozafindrabe)}
\label{sec:pA}

The interest of \pA\ collisions is based on the possibility  they open up to evaluate both the initial
and final-state effects on heavy-quarkonium production in cold nuclear
matter (CNM). Such baseline is mandatory to be able to draw
conclusions about any further effects due to QGP formation in \AaAa\
collisions. In the following, we will address a non-exhaustive 
list\footnote{ For instance, let us mention a potential further
significant effect due to charm-quark shadowing as computed in~\cite{Kopeliovich:2001ee}
in a light-cone Green function formalism.}
 of these CNM effects.

\paragraph{Initial-state effects}

Heavy-quarkonium production mainly proceeds through gluon fusion at
relativistic ion colliders. Therefore, the nuclear shadowing of
initial gluons has been extensively investigated
(see~\cite{nestorReviewShadowing} for a recent review), together with
its consequences on the charmonium production.

On the experimental side, the gluon nPDF\footnote{nPDF stands for the
parton density within a bound nucleon.} is loosely constrained at
small values of Bjorken-$x$:
\begin{itemize}
\item On the one hand, the processes used -- Deep-Inelastic scattering
      (DIS) and Drell-Yan -- are mostly sensitive to the quark and
      antiquark densities. Therefore, the gluon density is indirectly
      constrained, either via the nucleon-structure-function
      deviations from the Bjorken scaling, caused by gluon radiation,
      or via sum-rules (conservation of the nucleon momentum
      distributed among all partons).
\item On the other hand, there is no nuclear DIS data below $x\lesssim
      5 \,10^{-3}$ at perturbative values of the momentum transfer
      $Q^2 \gtrsim \Lambda_{\mathrm{QCD}}^2$, required for the
      validity of the DGLAP equations used to predict the evolution of
      nPDF with~$Q^2$.
\end{itemize}
As a result, the extracted parametrisations of the ratios nPDF/PDF
have large uncertainties at low-$x$: typically, for the gluon
shadowing in~Pb, the LO parametrisations
EKS98~\cite{Eskola:1998iy-Eskola:1998df} and
EPS08~\cite{Eskola:2008ca} give values of nPDF/PDF that differ by
about a factor of ten at $x = 10^{-4}$ and $Q^2 =
1.69~\mathrm{GeV}^2$. EPS08 notably includes additional constraints
from high-$P_T$ ($P_T \geq 2\mathrm{~GeV}/c$) hadron production
measured by the BRAHMS experiment~\cite{Arsene:2004ux} at forward
rapidities in \dAu\ collisions at RHIC top energy. These data mainly
probe gluons with $x \gtrsim 5 \, 10^{-4}$ in the gold nucleus and
suggest a much stronger shadowing. The discrepancy is even larger when
these LO parametrisations are compared to the NLO ones, either
nDSG~\cite{deFlorian:2003qf} or HKN07~\cite{Hirai:2007sx}. 
All these uncertainties preclude yet reliable predictions of 
heavy-quarkonium production at the LHC, dominated by low-$x$ 
gluons (see Section~\ref{sec:alice}).

A workaround to the use of these parametrisations can be found in
approaches that try to describe the shadowing in a formal way:
\begin{itemize}
\item The underlying physical mechanism is thought to be
      multiple-scattering (or multi-pomeron exchanges) with initial
      interactions between the pomerons, and the calculations are made
      in the Glauber-Gribov
      framework~\cite{GribovGlauberTh-Kaidalov:2007}. However, these
      models usually have a narrower validity range (limited to the
      low-$x$ region~\cite{Arneodo}) since they were designed to
      describe the coherence effects that lead to the depletion of the
      nuclear structure function. Indeed, at intermediate values of
      $x$, the amount of anti-shadowing appears to be smaller in these
      approaches than in the aforementioned parametrisations (for
      instance, see the comparisons reported in~\cite{FFR} for the
      $J/\psi$ shadowing in \dAu\ at RHIC).
\item The Colour Glass Condensate (CGC) is an effective theory which
      describes the behaviour of the small-$x$ components of the
      hadronic wavefunction in QCD (see~\cite{Iancu:2003xm} for a
      recent review). Hence, it can be used to study the high energy
      scattering in QCD, namely the initial stages of heavy-ion
      collisions. The theory is characterised by a saturation scale
      $Q^2_s$: at any $Q^2$ below this scale, the rapid rise of the
      gluon density at small-$x$ slows down to a logarithmic rate, due
      to a growing number of gluon-gluon fusions. It is of no doubt that
      \pA\ collisions at the LHC will provide crucial tests of the
      CGC framework, since they will allow to deeply probe the
      saturation region. At RHIC, the understanding of the initial
      effects on charmonium production in the CGC framework is still a
      work in progress. A first step was the
      calculation~\cite{Tuchin:2004ue} of the open charm production,
      which is found suppressed at forward rapidity at
      RHIC. Predictions for the $J/\psi$ are under way: for rapidities
      $y \geq 0$, a rather qualitative agreement with PHENIX \dAu\ data
      is obtained in~\cite{Kharzeev:2005zr}~\footnote{Note that the
      data to theory comparison made in this article should be updated
      with the newly published PHENIX \dAu\ results from the re-analysis
      described in~\cite{Adare:2007gn}.} for the $y$-dependence and
      for the centrality dependence~\footnote{This work is being
      extended in~\cite{NardiQM08} in order to describe the
      $y$-dependence of the peripheral \AuAu~collisions at RHIC. The
      first results seem quite promising.}.
\end{itemize}

To summarise, the amount of shadowing crucially depends on $x$. By
taking the $J/\psi$ transverse momentum~$P_T$ into account when
evaluating $x_1$, $x_2$ (and $Q^2$), the influence of~$P_T$ on the
shadowing is investigated in~\cite{FFR,FFLR} at RHIC. There is an
on-going debate on the way the $J/\psi$ acquires its~$P_T$: either
{\it(i)}~the initial gluons carry an intrinsic transverse momentum and
the latter is subsequently transferred to the $J/\psi$ ($2 \to
1$~process), or {\it(ii)} the $P_T$ comes from the emission of a
recoiling outgoing gluon ($2 \to 2$~process). For the latter process,
the authors consider the partonic cross-section given
in~\cite{schannelcut} which satisfactorily describes the $J/\psi$
$P_T$-spectrum down to $P_T\sim0$ at RHIC. On the average, initial
gluons involved in these processes originate from different
$x$-regions, hence resulting in quite different shadowing
effects~\cite{FFLR}. The present uncertainties on PHENIX \dAu\
data~\cite{Adare:2007gn} do not allow to discriminate these scenarios,
but forthcoming improvements will be obtained from the recent data
taking (with at least a factor of 30 in statistics).

Additional initial-state effects are the initial-partonic
multiple scattering and the related parton energy loss. It is believed
that the observed broadening of $\left<P_T^2\right>$ --~the mean value
of the $J/\psi$'s transverse momentum squared -- from \pp\ collisions
to \pA\ collisions (with increasing $A$) is due to such
multiple-scattering (the so-called ``Cronin effect''). This effect is
usually described as a random walk of the initial-projectile parton
within the target nucleus (see e.g.~\cite{Hufner:1988wz,Johnson:2000dm}): the resulting $\left<P_T^2\right>$
proportionally increases with the amount of scattering centers,
characterised by the length~$L$ of nuclear matter traversed. Simple
linear fits to $\left<P_T^2\right>$ vs $L$ can indeed account for the
\pA\ measurements done at SPS. All \pA\ and \AaAa\ results at SPS, but
the preliminary result reported by NA60 in \pA\ at $158
\mathrm{~AGeV}$, exhibit the same slope~\cite{Cortese:HP2008}. At
RHIC, the $\left<P_T^2\right>$ measured in \dAu~\cite{Adare:2007gn}
suffers from large uncertainties, but is compatible with a moderate
broadening. A linear fit done to all data points available at RHIC
(\pp, \dAu, {\it CuCu} and \AuAu) is reported
in~\cite{GranierdeCassagnac:2008ke}: the slope is compatible with zero
at mid~rapidity and with some broadening at forward
rapidity. Interestingly, within the large uncertainties, the slope
seen at forward rapidity at RHIC is compatible with the one at SPS
(the comparison can be made between the slopes quoted
in~\cite{GranierdeCassagnac:2008ke} and in the
slides~\cite{Scomparin:QM06}).

\paragraph{Final-state effects}

The so-called ``nuclear absorption'' has been extensively investigated for the charmonia. It reflects the break-up of correlated $c\bar{c}$ pairs due to inelastic scattering with the remaining nucleons from the incident cold nuclei. As we shall see below, the underlying mechanism is still unclear. Moreover, its physical picture may change with energy, as pointed in~\cite{Arsene:2007gx} and the transition would occur at RHIC energies:

\begin{itemize}
\item At low energy, there is a longitudinally-ordered scattering of the heavy-quark pair. It results in an 
attenuation factor $\exp{[-\sigma_{\mathrm{abs}}\,\rho_0 L(b)]}$, where $\sigma_{\mathrm{abs}}$ is the 
{\it effective} break-up cross-section, $\rho_0=0.17\mathrm{~nucleon}\,\mathrm{fm}^{-3}$ is the nuclear 
density and $L(b)$ is the length traversed by the heavy-quark pair in the nuclear matter at a given impact 
parameter~$b$. The break-up cross-section is not calculable from first-principle QCD\footnote{It is nevertheless
possible to extract its value from photoproduction data $\gamma N \to \psi N$ 
(see \eg~\cite{Hufner:1997jg,Hufner:2000jb}).}. Presently, it is a 
free parameter in the models, such as in~\cite{Adare:2007gn} where its value is obtained by fitting the 
data with a given nuclear shadowing model and an unknown additional absorption. Many considerations are 
hiding behind the effective value of $\sigma_{\mathrm{abs}}$. It has been 
argued~\cite{Kharzeev:1996yx,Vogt:1999dw,Vogt:2001ky,wohri:HP008} that the value of the break-up cross-section 
should depend both on the collision energy and on the colour state of the created $c\bar{c}$. With increasing 
collision energy, the pair will hadronise from inside to outside the nucleus. Since a hadronised $c\bar{c}$ is
 much more robust than the pre-resonance, the effective value of $\sigma_{\mathrm{abs}}$ will decrease with 
energy. A colour singlet pair has a smaller size and hence a shorter hadronisation time than a colour octet 
pair, so the corresponding break-up cross-section should be smaller. Moreover, if the direct $J/\psi$ and 
$\psi'$ are believed to be created perturbatively in the same $c\bar{c}$ octet state, then they should suffer 
the same amount of nuclear absorption, unless the hadronised states are also broken-up by the scattering off 
nucleons. A quite rich \pA\ program was developed at SPS, where the CNM effects can be described with the 
nuclear absorption only: the latest values from the NA50 experiment are~\cite{Alessandro:2006jt} 
$\sigma_{\mathrm{abs}}(J/\psi) = 4.2 \pm 0.5 \mathrm{~mb}$ and $\sigma_{\mathrm{abs}}(\psi') = 7.7 \pm 0.9 \mathrm{~mb}$. 
However, the quoted value of $\sigma_{\mathrm{abs}}(J/\psi)$ is for the total $J/\psi$ production: it accounts for 
the absorption of the direct $J/\psi$ as well as of the higher mass $c\bar{c}$ states that would have decayed 
in $J/\psi$ otherwise. It is also worth remembering that the shadowing effect was ``forgotten'' when evaluating
 the quoted values of $\sigma_{\mathrm{abs}}$. When properly taking into account the anti-shadowing at SPS as 
in~\cite{wohri:HP008}, the estimated break-up cross-sections are significantly larger (around $7\mathrm{~mb}$ 
for the $J/\psi$ and $11\mathrm{~mb}$ for the $\psi'$). At RHIC, the published values of the effective break-up 
cross-section in \dAu\ collisions~\cite{Adare:2007gn} are 
$\sigma_{\mathrm{abs}}(J/\psi) = 2.8^{+1.7}_{-1.4}(2.2^{+1.6}_{-1.5})\mathrm{~mb}$ for the two different nPDF/PDF 
parametrisations used to evaluate the shadowing. However, the shadowing used in~\cite{Adare:2007gn} is obtained
 without taking into account the $J/\psi$'s~$P_T$. When doing so, in the $2 \to 2$~process, the resulting 
shadowing is quite different, leading to a different value of the extracted break-up cross-section~\cite{FFLR}, 
which happens to be closer to the published NA50 value.
\item At high energy, the above space-time picture of the heavy-quarkonium production in \pA\ should not hold any more. As argued in~\cite{Arsene:2007gx,Capella:2007jv}, the heavy state in the projectile will rather undergo a coherent scattering off the target nucleons. The conventional treatment of nuclear absorption is not valid any more and the rigorous method reported in~\cite{Braun:1997qw} has to be used. As a result, $\sigma_{\mathrm{abs}}$ asymptotically tends to zero at high energies. The required threshold in $\sqrtsnn$ for this coherent scattering picture to be relevant is a function of $x_+$, the longitudinal momentum fraction of the heavy system. The transition happens at RHIC energy for the $J/\psi$ produced at mid rapidity. Consequently, the break-up cross-section is rapidity-dependent at RHIC. In~\cite{Arsene:2007gx,Capella:2007jv}, the authors use the method~\cite{Braun:1997qw} within a Glauber-Gribov description of shadowing to predict the CNM effects on the $J/\psi$ production at various $\sqrtsnn$, from $39~\mathrm{GeV}$ to the LHC energies. They report a fair agreement with E866, E772 and PHENIX $p\mathrm{(d)}A$ data, both for the rapidity-dependence and the $x_F$-dependence. 
\end{itemize}

\paragraph{The $J/\psi$ puzzle}

As can be seen from the above discussion, disentangling the various CNM effects at work for the $J/\psi$ production in \pA\ is not an easy task. So, in order to set a proper baseline for the QGP search, how can we safely extrapolate the CNM effects from \pA\ to \AaAa\ ? The current solution adopted by PHENIX~\cite{Adare:2007gn} is to use the data-driven method inspired by~\cite{GranierDeCassagnac:2007aj} to predict the expected CNM effects in \AuAu~from the available \dAu\ data. This approach has the advantage of a proper propagation of the experimental errors. The quite annoying drawback is that it is unable to derive any specie- nor energy- dependent extrapolation of the CNM effects. 

To improve this situation, we clearly need more and much precise data. At RHIC, the analysis of the latest high statistics \dAu\ run is well under way. At the LHC, a rich \pA\ program will be crucial.

In the meantime, a key test to the models would be to reproduce the observed scaling of the $\alpha$ exponent in $\sigma_{p\mathrm{(d)}A} = \sigma_{pp}A^{\alpha}$, with $x_F$ for the $J/\psi$ production in \pA\ at various $\sqrtsnn$. At least three different approaches tried to cope with this observation. A first tentative is done in~\cite{Vogt:1999dw,Vogt:2001ky}, with a combination of five different types of CNM effects. More recently, with the help of the CGC approach and a simple (single-valued) $\sigma_{\mathrm{abs}}$, the authors of~\cite{Kharzeev:2005zr} obtained a result with a reasonable agreement to the observed scaling. In the previous paragraph, we already mentioned the positive result achieved in~\cite{Arsene:2007gx} by the use of a coherent scattering picture of the nuclear absorption and a Gribov-Glauber framework to describe the shadowing.

\subsection{\AaAa\ collisions  \protect\small (by Joseph Cugnon)}
\label{sec:issues-AA}

The interest of heavy-quarkonium production in heavy-ion collisions is twofold. 
First, these collisions may offer different mechanisms of quarkonium production
 compared to \pp\ collisions. Second, and more importantly, this production can 
be used as a sensitive probe to the supposedly formed quark-gluon plasma in the course of these collisions. There is a general consensus that, in the SPS and 
RHIC conditions, the quarkonium states (mainly \jpsi) are formed during the 
first nucleon-nucleon ($NN$) collisions as in free space and that they are immersed
 afterwards in the  quark-gluon plasma, where they can be dissolved due the 
screening of the quark-antiquark interaction in the plasma. In these conditions 
the plasma is rather cold, few charmed quarks are thermally produced and the 
final \jpsi~yield results from the absorption of the early produced charmonium 
states by the plasma. The situation will be somehow different at the LHC. The 
temperature of the plasma will be substantially larger, free charmed quarks 
will be more numerous. Furthermore, charm production will result also from 
the decay of bottom, which will be produced sizeably at $\sqrtsnn$~=~5.5~TeV. 
In addition to the usual formation in first $NN$ collisions, charmonium states 
will be produced also through the decay of $B$-mesons and through the fusion of 
$c\bar c$ pairs, either thermally produced directly or through the decays of free 
$b$-quarks (the so-called regeneration process).  In relation to the use of quarkonia 
as probes of the plasma, these additional production processes should be clarified 
both experimentally and theoretically. First the side-feeding in the propagation 
of excited states in the plasma should be evaluated correctly. It should be reminded 
that the side-feeding of \jpsi~is not yet satisfactorily clarified in the previous 
experiments~\cite{GR08}. As indicated in Section~\ref{sec:alice}, this problem may be circumvented 
by using the $\Upsilon$ family as probes of the plasma. Nevertheless, 
the reliable evaluation of the feeding of \jpsi~by higher mass resonances in 
the plasma seems to be a necessary step on the analysis of future experiments.
 Furthermore, if one is interested in the energy or $P_T$ spectrum of produced quarkonia, 
the study of the propagation of heavy quarks is highly desired. Progress have
 been made recently~\cite{GO08}, with the use of Fokker-Planck equations (in
 relation to jet quenching), but the application of this approach is just 
beginning~\cite{PE08}.

The discussion of our present understanding of quarkonium production in terms of 
known properties of these objects and the supposed properties of the plasma is 
postponed to Section~\ref{sec:bridging-the-gap}. It probably requires a reliable description of the 
space-time evolution of the plasma and a good understanding of the interaction 
between a quarkonium and the surrounding medium. In the past, this problem has 
been tentatively circumvented by focusing on variables that are hopefully not 
sensitive to the detail of this evolution. The most popular example is provided 
by the famous plots of $R_{AA}$ versus $L$, the estimated average path of the 
quarkonium inside the plasma and admittedly a very crude variable. This has 
obvious limitations and does not presently provide a coherent view of the 
existing experimental data~\cite{KH08}. Therefore, an important effort in order 
to improve the theoretical description of the evolving plasma is certainly required. 

Concerning the coupling of the quarkonia to the plasma, new lattice-QCD
developments have taken place recently. There are more and more
indications that quarkonia may survive in the plasma up to
temperatures of the order of two times the critical
temperature~$T_c$~\cite{KA}. Whether the origin of this property
results from unsuspected aspects of the screening properties of the
plasma~\cite{MO08} or from other sources is beyond the scope of this
overview.  As a consequence, the situation at the LHC will be more
comfortable than before, since the temperature of the plasma will be
higher. In any case, a good theoretical understanding of the coupling
is necessary. It is not granted that this coupling is describable in
terms of cross sections for quarkonium-gluon collisions. If it is the
case, recent progresses have been made in this field~\cite{AR05}, although the
variation of the cross section with energy seems to pose
problems~\cite{GR08}.

Finally, the diagnostic of the plasma will be possible only if the other 
probes are also well understood: high-$P_T$ photons, dileptons 
(coming from low mass resonances), etc. The first one is sometimes 
considered as the ideal probe. However, the present situation is not
 clear. There are inconsistencies between \AuAu~and {\it CuCu} data~\cite{GR08}. 
The second one seems to be more distorted than  mesonic resonances by
medium effects. Further theoretical work is necessary.

\section{Recent theoretical advances}

\subsection{QCD corrections  \protect\small (by Jean-Philippe Lansberg)}
\label{sec:QCD-corrections}

Recently, substantial progress has been achieved in the computation
of higher-order QCD corrections to the hard amplitudes of 
quarkonium-production processes. The first NLO calculation to date was centered
on unpolarised photoproduction of $\psi$~\cite{Kramer:1995nb} via 
a colour-singlet (CS) state~\cite{CSM_hadron} (that is, LO in $v$ for
NRQCD~\cite{Bodwin:1994jh}) more than ten years ago now. Later on, NLO
corrections were computed for direct $\gamma\gamma$
collisions~\cite{Klasen:2004az,Klasen:2004tz} where it had been
shown~\cite{Klasen:2001cu} previously that the CS contribution alone 
was not able to correctly reproduce the measured rates by DELPHI~\cite{Abdallah:2003du}.

At the LHC and the Tevatron, $\psi$ and $\Upsilon$ production proceeds
most uniquely via gluon-fusion process. The corresponding cross
section at NLO ($\alpha_S^4$ for hadroproduction processes) are
significantly more complicated to compute and became only available
one year ago~\cite{Campbell:2007ws,Artoisenet:2007xi}.  Those results
were recently confirmed in~\cite{Gong:2008sn,Gong:2008hk}. In the
latter papers, the polarisation information was kept and the
observable $\alpha$ was also computed. It is important to stress that 
for $\psi$ and $\Upsilon$ production the CS yields predicted at 
the NLO accuracy are still clearly below the experimental data. 
In this respect, the predictions for the polarisation at this order 
cannot be usefully compared to the data.

Aside from hadroproduction regime, NLO corrections have also recently
been computed for two $J/\psi$-production observables at the
$B$-factories: $J/\psi + c \bar c$~\cite{Zhang:2006ay}
and $J/\psi+\eta_c$~\cite{Zhang:2005cha}.

The common feature of these calculations is the significant size of
the NLO corrections, in particular for large transverse momenta  $P_T$ of the
quarkonia when this information is available. In $\gamma p$ an \pp~collisions, 
QCD corrections to the CS production open new channels with
a different behaviour in $P_T$ which indeed raise substantially the cross
section in the large-$P_T$ region. In general, the
CS prediction can thus be brought considerably closer to
the data, although agreement is only reached at NLO in the photoproduction
case~\cite{Kramer:1995nb}. As of today, only the full colour-octet (CO) 
contributions to direct $\gamma\gamma$ collisions have been evaluated
at NLO for $P_T>0$~\cite{Klasen:2004az,Klasen:2004tz}. Very recently, 
CO contributions from $S$ waves ($^1S_0^{[8]}$ and $^3S_1^{[8]}$) have
become available~\cite{Gong:2008ft} for hadroproduction, but their impact
on phenomenology has not been fully studied yet. Since NLO corrections do 
not affect significantly the $P_T$ dependence, a quick assessment can be
obtained via the $K$ factors, the ratios of NLO to LO predictions. 
The $K$ factors of the cross section  at
the Tevatron are about 1.2 for the $^1S_0^{[8]}$ state and 1.1 for the
$^3S_1^{[8]}$ one (at the LHC, they are both about 0.8).  This 
entails that the value of the CO Long Distance Matrix Elements (LDMEs) fit to the Tevatron data at LO
$\langle O \big( ^3S_1^{[8]}\big) \rangle \simeq 0.0012$~GeV$^3$ and
$\langle O \big( ^1S_0^{[8]}\big) \rangle\simeq 0.0045$
GeV$^3$~\cite{Kramer:2001hh} would be at most reduced by
15\%. In this respect,  the NLO corrections to the octets do not improve
the universality of the matrix elements when the idea of the dominance of
the CO transitions is confronted to the data on photoproduction from HERA.

In~\cite{Gong:2008ft}, the authors made a fit of the CO LDMEs on
prompt data which is not easily comparable to the previous
works since it includes feeddown from $\psi'$ and $\chi_c$.  It is
however worthwhile to note that they had {\it to abandon} [in the fit]
{\it the experimental data with $P_T < 6$~GeV/c, since it is not possible
to obtain a satisfactory $P_T$ distribution in terms of a unique
$\langle O^H_n \rangle$ value}. This emphasises the need for more work
dedicated to the description of the low-$P_T$ region.  Last but not
least,  the polarisation from CO 
transitions appears not to be modified at NLO with respect to LO result,
thus confirming the flagrant discrepancy between the NRQCD predictions for
the polarisation of the $J/\psi$ and the experimental measurements from the CDF
collaboration~\cite{Abulencia:2007us}.

On the bottomonium side, the situation seems less problematic.  We
have now at our disposal the CS cross
section at NLO including a dominant subset of NNLO corrections at
$\alpha_S^5$ (namely the associated production
 with 3 light partons) for inclusive $\Upsilon$
 hadroproduction~\cite{NNLO} at mid and large $P_T$.  The rate
 obtained by including only the CS channels
are in substantial {\it agreement} with the experimental measurements
of the cross section from the
Tevatron~\cite{Acosta:2001gv,Abazov:2005yc} .  Concerning the
polarisation, the {\it direct} yield is predicted to be mostly
longitudinal. The experimental data being centered
 on {\it prompt} yield~\cite{Acosta:2001gv,D0:2008za}, we would need
 first to gain some insights on NLO corrections
to $P$-wave production at $P_T>0$ to draw further conclusions. Yet,
since the yield from $P$-wave feeddown is likely to give transversely
polarised $\ups$, the trend is more than encouraging.

\subsection{Automated  generation of quarkonium
amplitudes in NRQCD  \protect\small (by Pierre Artoisenet)}

The computation of heavy-quarkonium cross section within Non-Relativistic QCD~\cite{Bodwin:1994jh}
takes advantage of the \textit{small} relative velocity $v$ inside the 
quarkonium state to factorise in a consistent way perturbative high-energy
effects (linked to the heavy-quark-pair production) from non-perturbative
low-energy effects (linked to the evolution of the heavy-quark pair into a quarkonium
state). This factorisation is performed order by order in $\alpha_S$ (the strong
coupling constant) and in $v$, and is controlled by a factorisation scale $\Lambda$.
As a result, differential cross sections read 

\begin{equation}
d \sigma (\mathcal{Q}) = \sum_{n} d\hat{\sigma}_\Lambda \left( Q \bar Q(n) \right) 
\langle \mathcal{O}^\mathcal{Q}(n)\rangle_\Lambda
\label{NRQCD_crossX}
\end{equation}
 where $n$ specifies the quantum numbers of the intermediate heavy-quark pair~$Q \bar Q$. 
The factors $d\hat{\sigma}_\Lambda \left( Q \bar Q(n) \right)$ are called
the short-distance  coefficients, and can be computed perturbatively in $\alpha_S$
for a given process. The Long-Distance Matrix Elements (LDMEs)
$\langle \mathcal{O}^\mathcal{Q}(n)\rangle_\Lambda$ encode the soft evolution
of the heavy-quark pair into a quarkonium state. They are universal, i.e., they do not
depend on the details of the creation of the heavy-quark pair.

In order to predict the cross section for a given process with a quarkonium in the final
state, one has to compute the short distance coefficients at a given accuracy in $\alpha_S$
and in $v$ (which limits the number of transitions $n$ in the sum in Eq. (\ref{NRQCD_crossX})).
Even for the tree-level amplitudes, a calculation by hand may be lengthy and in general 
error-prone, depending on the parton multiplicity in the final state. For the purpose of a fast, 
easy-to-handle and reliable computation of tree-level quarkonium amplitudes, a new MadGraph-based
implementation~\cite{Artoisenet:2007qm} (MadOnia) has been developed recently. 
The user may require an $S$- or a $P$-wave, colour-singlet (CS) or colour octet (CO) intermediate state,
for any process attainable in MadGraph for the open-quark production.
The code then generates automatically the related squared amplitude at leading order
in $v$, which can then be interfaced with a phase-space generator to obtain
the short-distance cross section. For $S$-wave state production, the algorithm can also be
extended to compute relativistic corrections to the Born-level cross section.
For $^3S_1$ states, the decay into leptons can be included, thus giving 
an easy handle onto polarisation studies. Another capability of the code is the 
generation of  amplitudes  involving heavy quarkonium with mixed flavors, such as the $B_c$.  

Several applications of the code have already been reported
~\cite{Artoisenet:2007qm,Artoisenet:2007xi,Artoisenet:2008tc,NNLO}. 
One example is the analysis of the associated production of
a $J/\psi$ or a $\Upsilon$ plus a heavy-quark pair of the same 
flavor at the Tevatron. The prediction of differential cross sections as well 
as polarisation observables, both for CS and CO transitions,
is straightforward. One can directly show, for example, that
the CS fragmentation approximation, which 
was used to estimate the $J/\psi +c \bar c$ production at the Tevatron,
appears to underestimate the yield by a large factor  in the region 
$P_T<20$~GeV/$c$. 

Beside offering the possibility to check very efficiently the tree-level contribution
of a large set of quarkonium production processes, the code is also embedded
in a larger project aimed to serve the connection between theory and experiments.
At the LHC, $J/\psi$ or $\Upsilon$ hadro-production followed by their leptonic decay
will offer a very clean signature to calibrate the detectors as well as to probe new 
physics effects. Given the large number of such events, one can even hope to reconstruct
more exclusive final states, such as $\Upsilon+2$ $b$-jets, and hence enlarge the number 
of measured observables to be compared with theoretical predictions. On the one
hand, such measurements  rely on theoretical assumptions
to establish the criteria for the selection of the signal and to estimate the cut efficiency. 
On the other hand, theoretical predictions make use of the existing 
experimental data to quantify non-perturbative low-energy effects,
including the values of the  LDMEs
in the case of quarkonium production.

The flow of information between theory and experiment is made easier
by the use of Monte Carlo tools. The new generation of these tools
operate in two steps.  First they produce parton-level events
according to a hard scattering probability which can be computed
perturbatively in $\alpha_S$. The events are then passed through a
code that generates the parton shower and turns the partons into
hadrons. Eventually one can use a detector simulator to smear the
information according to the resolution of the detector, such that
events are as close as possible to real data.  For quarkonium
production within NRQCD, the relative abundance of parton-level events
is controlled by the short distance coefficients appearing in
Eq. (\ref{NRQCD_crossX}), which can be computed by MadOnia. The
algorithm is being currently promoted to an event generator. Spin
correlation and colour flow information are kept, such that the
unweighted events are ready for parton shower and hadronisation. With
such a generator at hand, a large set of studies will become available.

\subsection{Other theoretical advances \protect\small (by Jean-Philippe Lansberg)}
\label{sec:th-adv-in-pp}

On top of the theoretical advances mentioned above, several  
interesting theoretical results have been obtained in recent years. 
Let us review some of the most significant ones briefly.

Last year, Collins and Qiu~\cite{Collins:2007nk} showed that in
general the $k_T$-factorisation theorem does not hold in production of
high-transverse-momentum particles in hadron-collision processes, and
therefore also for $\psi$ and $\Upsilon$. This is unfortunate since
many studies had been carried out successfully using $k_T$-factorisation 
(see references in section 3.3
of~\cite{Lansberg:2006dh}), predicting mostly longitudinal yields and
smaller CO LDMEs, in better agreement with the idea of LDME
universality.

On the side of NRQCD, Nayak, Qiu and Sterman provided an up-to-date 
proof~\cite{Nayak1} of NRQCD factorisation holding true at any order
in $v$ in the gluon-fragmentation channels. They showed that improved 
definitions of NRQCD matrix elements were to be used, but that
this was not to affect phenomenological studies.

Besides, the $c$- and $b$-fragmentation approximation was shown to fail 
for the  $P_T$ ranges accessible in experiments for 
quarkonium hadroproduction~\cite{Artoisenet:2007xi}.
By studying the entire set of diagrams contributing to $\psi$ and 
$\Upsilon$ production in association with a heavy-quark pair of the same
flavour, it was shown that the full 
contribution was significantly above (typically of a factor of 3)
that obtained in the fragmentation approximation. The latter 
holds (at 10\% accuracy, say) only at very large
 $P_T$: $P_T \gtrsim 60$~GeV/$c$ for $\psi$ and $P_T \gtrsim 100$~GeV/$c$ for 
$\Upsilon$.  Note that the same observation was previously 
made for the process $\gamma \gamma \to J/\psi c 
\bar{c}$~\cite{Qiao:2003ba} and also for the  $B_c^*$ hadroproduction,
for which it was noticed that the fragmentation approximation was not
reliable at the Tevatron~\cite{Chang:1995,Berezhnoy:1996ks}.

Moreover, still in double-heavy-quark-pair production, the notion of 
colour-transfer enhancement was introduced by Nayak, Qiu and 
Sterman~\cite{Nayak:2007mb}. If three out of the four heavy quarks 
are produced  with similar velocities, then there is the possibility that 
colour exchange within this 3-quark system could turn a CO
configurations into CS ones, thus effectively increase the rate of production of 
CS pairs. They finally discussed the introduction of specific 
new 3-quark operators of NRQCD necessary to deal with such an issue.

\section{Perspectives for some (new) observables}
\label{sec:new-obs}

Standard quarkonium measurements with general purpose detectors at the 
LHC are generally related to kinematical distributions of the quarkonium 
decay products, such as differential cross section and polarisation 
measurements, and focus on decays into muons. Although these provide 
useful information, it is important to investigate the use of additional 
observables. In particular, it could be helpful to take into account not 
only the kinematics of the quarkonium itself, but also that of particles 
produced in association, or their nature as for instance in 
the study the production of quarkonia in association with a heavy-quark pair.

\subsection{Hadronic activity around the quarkonium  \\ \protect\small (by A.C. Kraan)}

Here, we investigate observables that are sensitive to the hadronic activity 
directly around the produced quarkonium~\cite{mycontribution}. This allows to extract 
information about the radiation emitted off the coloured heavy-quark pair 
during the production, and thereby about its production mechanism itself. To study the 
sensitivity of a typical multi-purpose LHC detector, we generated $J/\psi$ 
and $\Upsilon$ events in PYTHIA 8~\cite{pythia8} in four production toy models: colour-singlet 
and three colour-octet models with a varying amount of shower evolution 
of the coloured $Q\bar{Q}$-state. 

Because at lower \ptjpsi, shower activity is small in general, differences 
manifest themselves only at higher values of \ptjpsi, about 20~GeV/$c$. It 
must be noted that the energy of the surrounding particles associated with
 \jpsi-production is small (order~GeV), and as such it is not a priori clear
 whether there is any sensitivity at all over the underlying-event background. Also, a careful understanding of 
the detector is necessary, so we would not recommend this analysis to be done with early LHC data.

\begin{figure}[h!]
\includegraphics[width=6.5cm]{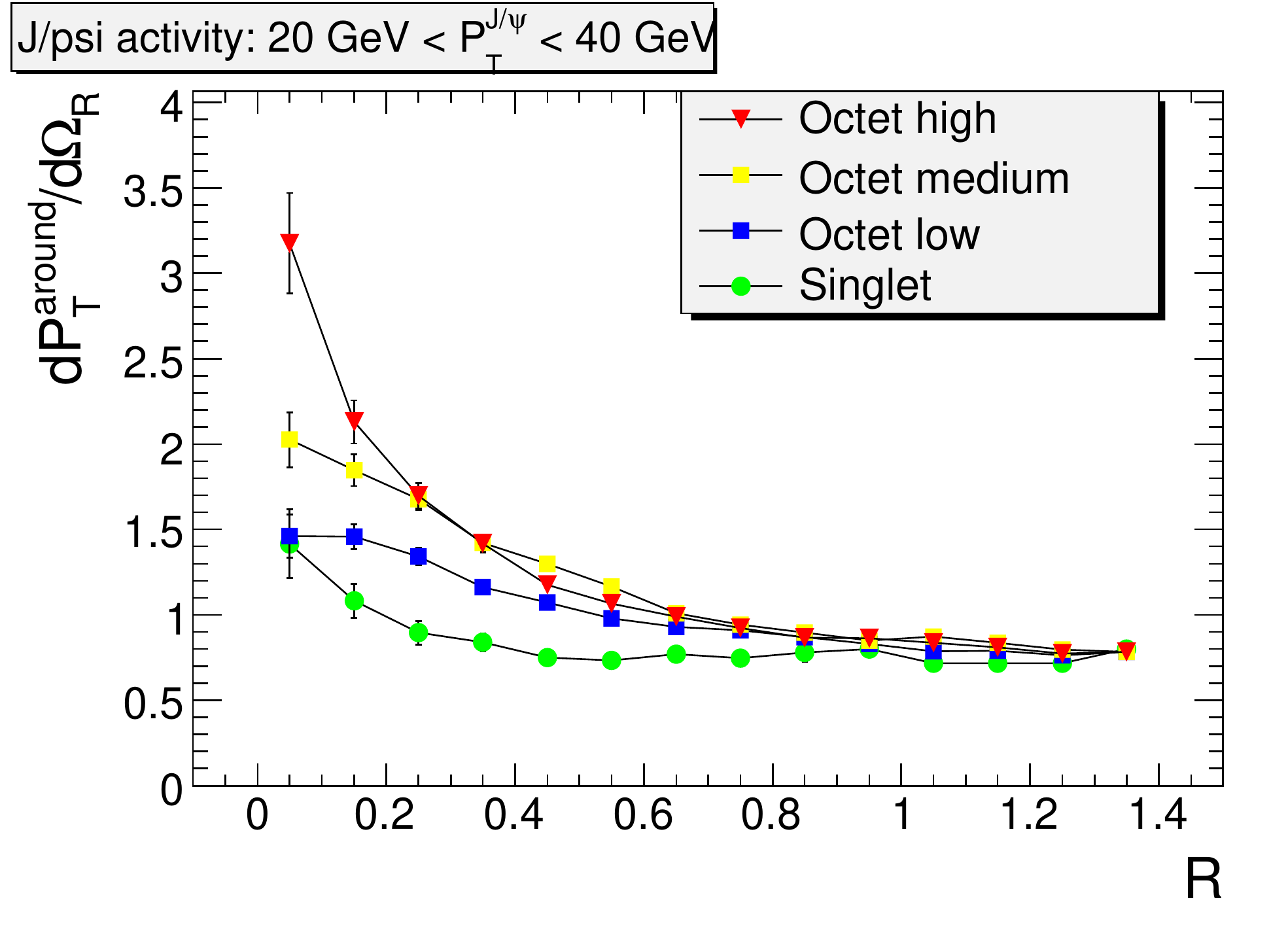}\hspace*{-0.4cm}
\includegraphics[width=6.5cm]{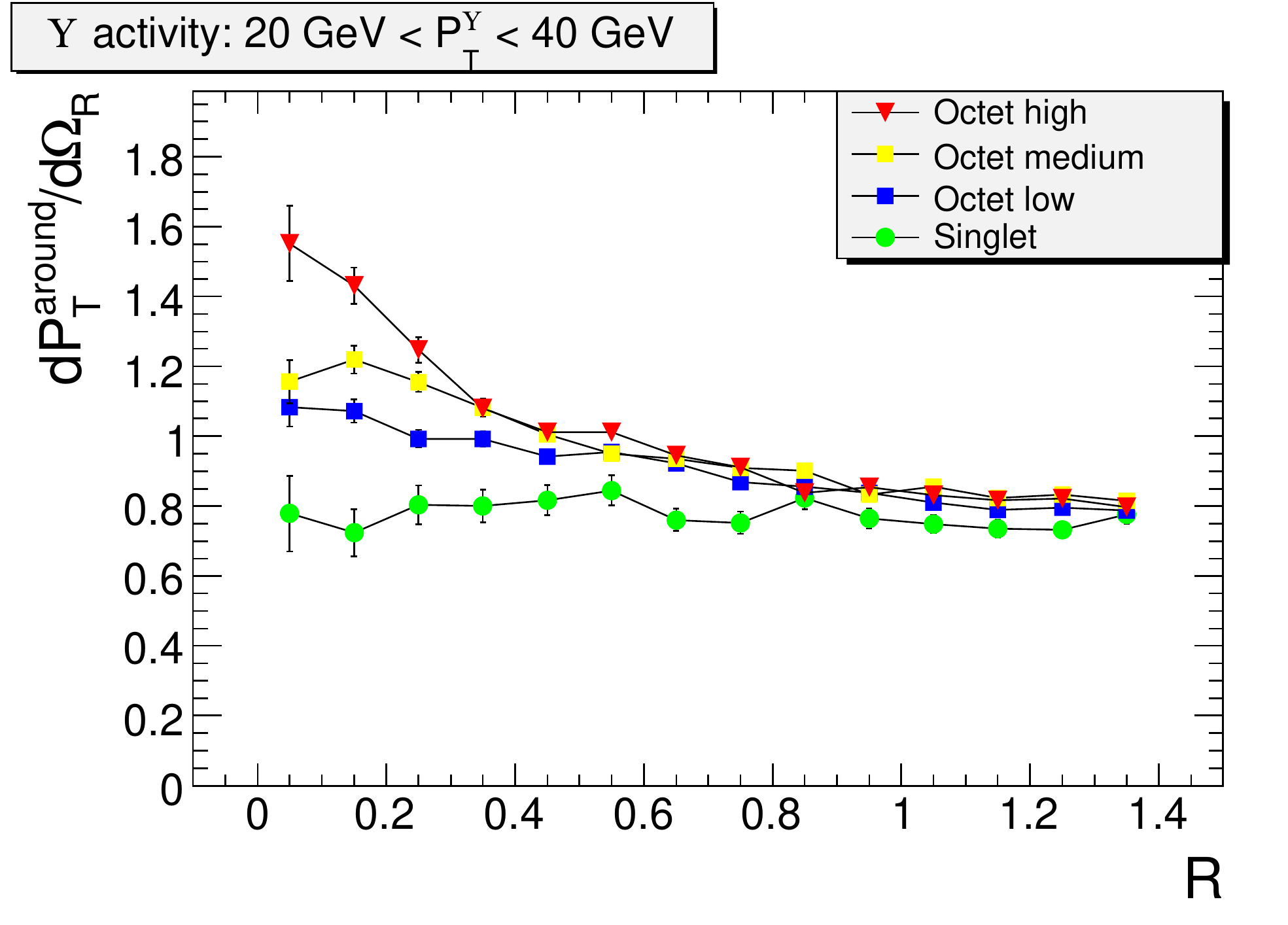}
\caption{\footnotesize Variable $d P_{T}^{around}(R)/d\Omega_R$ 
at reconstruction level for $J/\psi$ (left) and $\Upsilon(1S)$ (right) events. \label{z}}
\end{figure}

In Fig.~\ref{z} (left) we display the transverse momentum density 
d$P_T$/d$\Omega_R$ for \jpsi's between 20 and 40~GeV/$c$ after reconstruction, 
in a cone around the \jpsi~of certain size $R=\sqrt{(\Delta\eta)^2+(\Delta\phi)^2}$, where

\begin{equation}
\frac{\rm d P_{T}^{around}(R)}{\rm d\Omega_R}=\frac{\rm P_{T}^{around}
(\rm  R+d R/2)-P_{T}^{around}(R-dR/2)}{\rm \pi[( R+dR/2)^2-(R-dR/2)^2]}.
\end{equation}
 Here $\rm P_{T}^{around}(R)$ is the sum of the transverse momentum of all 
charged particles (with $P_T>0.9$~GeV/$c$) inside the cone of size $R$. In
 Fig.~\ref{z} (right) we display this variable for prompt $\Upsilon(1S)$-events. 
For $\Upsilon$-events the activity is lower than that for $J/\psi$-events
 because the $b\bar{b}$ is heavier than the $c\bar{c}$ state, and thus has 
a smaller shower evolution.

For prompt $J/\psi$'s, the main background comes from non-prompt 
\jpsi~production. Whether one can subtract the background component from 
data in order to obtain the hadronic activity related to prompt production 
remains to be seen. For $\Upsilon$'s, the situation is easier: the 
background can be studied accurately by studying the events in the side-bands. 

\subsection{Associated production channels  \protect\small (by Jean-Philippe 
Lansberg)}

Another very valuable observable, likely to test the many production models 
available~\cite{Brambilla,Lansberg:2006dh}, is the study of 
associated production channels, first in \pp\ collisions, then in \pA\ and 
\AaAa. By associated production channels, we refer to $\psi + c \bar c$ and 
$\Upsilon + b \bar b$.

A first motivation for such studies is simple: similar studies carried at 
$B$-factories showed an amazingly large fraction of $J/\psi$ production 
in association with  another $c \bar c$ pair. Indeed, 
Belle collaboration first found~\cite{Abe:2002rb} $\frac{\sigma\,(e^+ e^- 
\to J/\psi +c \bar c)}{\sigma\,(e^+ e^- \to J/\psi +X)}$
to be $0.59^{-0.13}_{+0.15}\pm 0.12$. Thereafter, the analysis was 
improved and they obtained~\cite{Uglov:2004xa} 

\eqs{\frac{\sigma\,(e^+ e^- \to J/\psi +c \bar c)}{\sigma\,(e^+ e^- \to 
J/\psi +X)}&= 0.82 \pm 0.15 \pm 0.14,\\
&> 0.48 \hbox{ at 95\% CL}.}
 
Whether or not such a high fraction holds for hadroproduction as well, is 
a question which remains unanswered. Analyses at the Tevatron 
(CDF and $D\emptyset$) and at RHIC (PHENIX and STAR) are already possible. 
As computed in~\cite{Artoisenet:2007xi} 
for the RUN2 at the Tevatron at
$\sqrt{s}=1.96$ TeV, the integrated cross-section are significant (see~\cf{fig:associated-sig} for the \jpsi):

\eqs{\sigma(J/\psi +c \bar c) \times {\cal B} (\ell^+\ell^-)& \simeq 1~ 
\hbox{nb}\\
\sigma(\Upsilon +b \bar b)\times {\cal B} (\ell^+\ell^-)& \simeq  1~ 
\hbox{pb}}

\begin{figure}[h!]
\includegraphics[width=8cm]{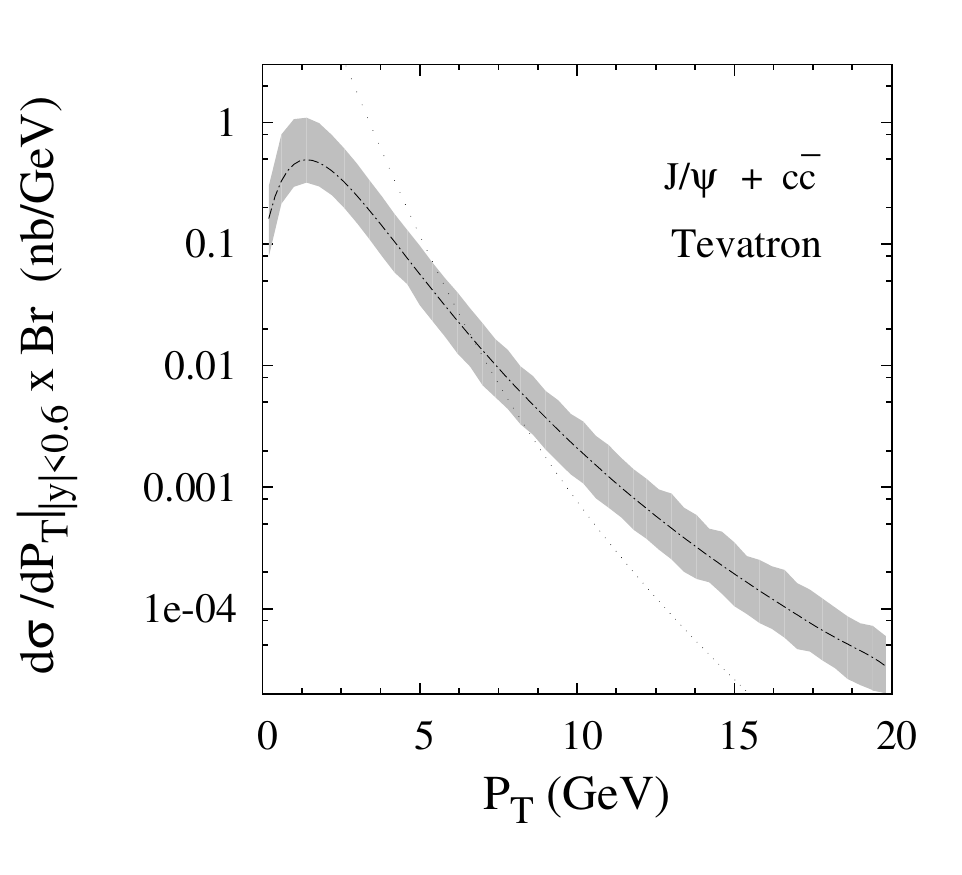}
\caption{Differential cross section for $pp \to J/\psi +c \bar c$ as function of the $J/\psi$ transverse 
momentum $P_T$ at the Tevatron ($\sqrt{s}=1.96$ TeV).} \label{fig:associated-sig}
\end{figure} 

Without taking into account any modifications of the CO LDMEs 
induced by  the QCD corrections mentioned in Section \ref{sec:QCD-corrections}, 
the integrated cross sections were found in~\cite{Artoisenet:2008tc} to be 
dominated by the CS part, similarly to the differential cross 
section in $P_T$ up to at least 5~GeV/$c$ for $\psi$ and 10~GeV/$c$ for $\Upsilon$. In other words, such
observables can be thought of as a test of the CS contribution, 
for the first time since the idea that CO transitions would be the dominant 
mechanism responsible for quarkonium production at high transverse momentum.
If the effect of CO transitions is confirmed to be negigible for the $\Upsilon$, 
the $\Upsilon$ produced in association with a $b \bar b$ 
pair are predicted to be strictly unpolarised, for any $P_T$ (see \cf{fig:associated-pol}).

\begin{figure}[h!]
\includegraphics[width=10cm]{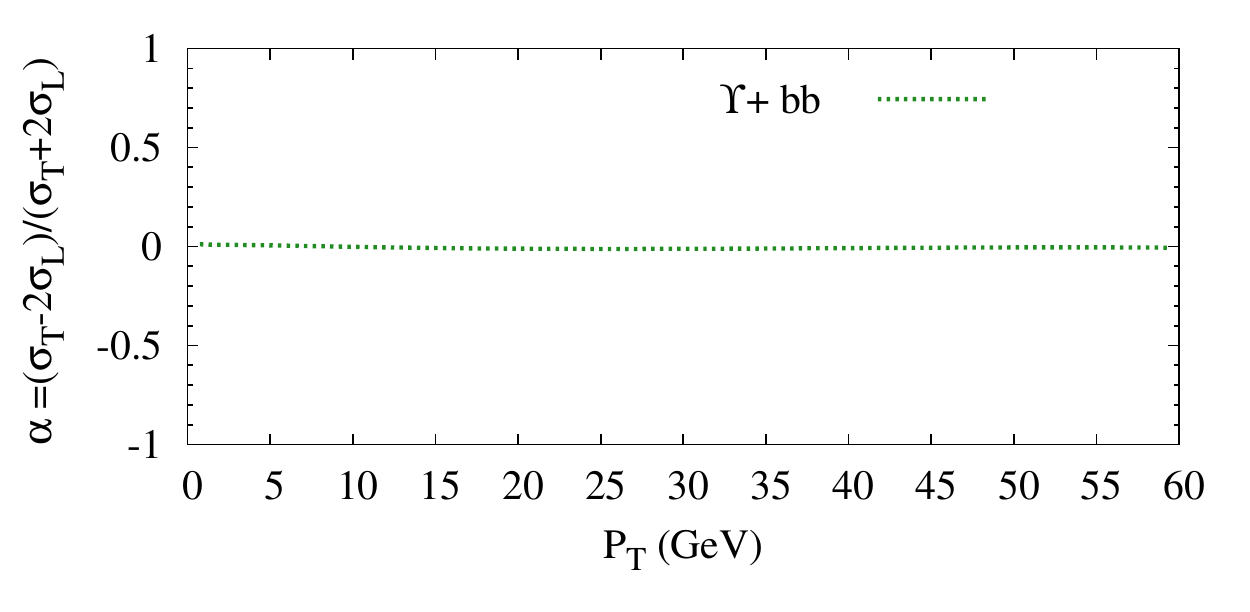}
\caption{Polarisation of an $\Upsilon$ produced in association with a 
$b\bar b$ pair at the Tevatron 
for $\sqrt{s}=1.96$ TeV for $|y|\leq 0.6$.} \label{fig:associated-pol}
\end{figure} 

Beside the property of discriminating between the CO and the CS 
transitions, the yield of $\psi$ in association with $c\bar c$ should be a priori less
sensitive to the $\chi_c$ feeddown and $B$ feeddown (and $\Upsilon$ with $b\bar b$ 
insensitive to the $\chi_b$ feeddown). Indeed, being suppressed by the relative velocity,
the $P$-wave yield is expected to be smaller than the CS 
$S$-waves\footnote{To be complete, let us mention the possibility to produce $\chi_c+c\bar c$
 via the process $gg\to gg$ for which the two final-state gluons split into a $c \bar c$
pair, one of them habronising into a  $\chi_c$ via the CO mechanism. 
This contribution is certainly suppressed up to $P_T \simeq 20$ GeV. For larger $P_T$,
a dedicated calculation is needed. However, this mechansim would be very easily disentangled 
from the CS contributions since both $c$ quarks are necessarily emitted back to back
to the  $\chi_c$ and thus to the $J/\psi$.
}.
 Note that the situation is completely at variance with the inclusive case 
for which it is nearly always easier to produce a $P$-wave
than a $\psi$ since this requires one less gluon attached to the heavy-quark 
loop.  Concerning the $\psi$ feeddown from $B$, the same is expected to occur: 
there are only prices to pay to produce a $\psi$ via a $B$ with 
a $c \bar c$ (heavier quark mass and then decay into a $\psi$) while no gain in the 
$P_T$ dependence since both $B(\to \psi X) + c\bar c $ and $\psi +c \bar c$ 
cross sections scale like $P_T^{-4}$.

Let us also mention that associated production has also been studied in 
direct $\gamma\gamma$ collisions in Ultra-peripheral collision 
(UPC)~\cite{Klasen:2008mh}. At least for direct
$\gamma\gamma$ collisions, associated production is the dominant 
contribution to the inclusive rate for $P_T \geq 2$~GeV/$c$.

To conclude, let us mention that studies can be carried on by
detecting either the ``near'' or ``away'' heavy-quark with respect to
the quarkonia. There are of course different way to detect the $D$,
$B$, or a $b$-jet, ranging from the use of a displaced vertex 
to the detection of their decay in $e$ or $\mu$. This has 
to be considered by also taking into account the different backgrounds.
In any case, we hope that such measurements would provide with clear
information on the mechanism at work in quarkonium production.

\subsection{Exclusive quarkonium photoproduction in proton 
and nucleus colliders \protect\small (by David d'Enterria)}

\begin{figure}[htpb]
  \centerline{\includegraphics[width=0.3\columnwidth]{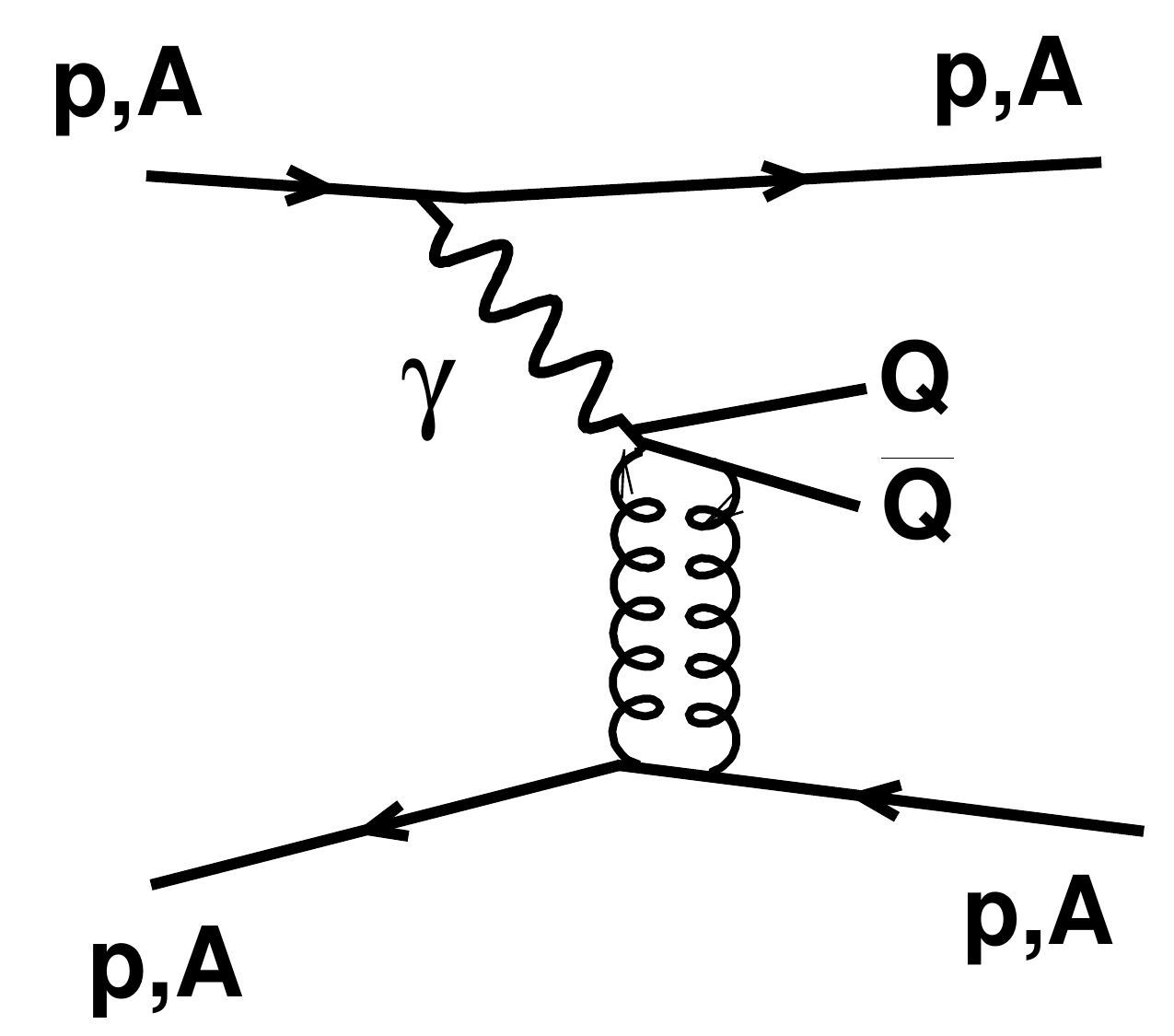}}
 \caption{Schematic diagram for exclusive diffractive quarkonium photoproduction in $\gamma A$, $\gamma p$ collisions
collisions.\label{fig:qqbar_diag}}
\end{figure}

A significant fraction of proton-proton and ion-ion collisions at collider energies involve 
``ultraperipheral''  electromagnetic interactions characterised (in the Weizs\"acker-William 
equivalent-photon-approximation~\cite{WW}) by the exchange of a quasi-real photon. Exclusive quarkonium 
photoproduction at proton or nucleus colliders -- i.e. processes of the type $\gamma\, h\rightarrow V\,h$ 
where $V=\jpsi,\ups$ and the hadron $h$ (which can be a proton or a nucleus $A$) remains intact -- 
has been measured at RHIC~\cite{dde_qm05} and the Tevatron~\cite{pinfold08} and will be measured 
at the LHC in both \pp~\cite{jhollar08} and \pPb\ and \PbPb~\cite{UPCreport} collisions. 
Exclusive $\qqbar$ photoproduction offers an attractive opportunity to constrain 
the low-$x$ gluon density at moderate virtualities, since in such processes the gluon 
couples {\em directly} to the $c$ or $b$ quarks (see Fig.~\ref{fig:qqbar_diag}) and the cross section is proportional to 
the gluon density {\em squared} (see~\cite{teubner07} and refs. therein). The mass of 
the $Q\bar Q$ vector meson introduces a relatively large scale, amenable to a perturbative 
QCD (pQCD) treatment. In the case of nuclei, the information provided by such processes 
is especially important since the gluon density is very poorly known at low-$x$ and there 
are not many experimental handles to measure it in a ``clean'' environment~\cite{dde_photon07}.\\

\begin{figure}[htpb]
\centerline{\includegraphics[width=5cm,height=4.5cm]{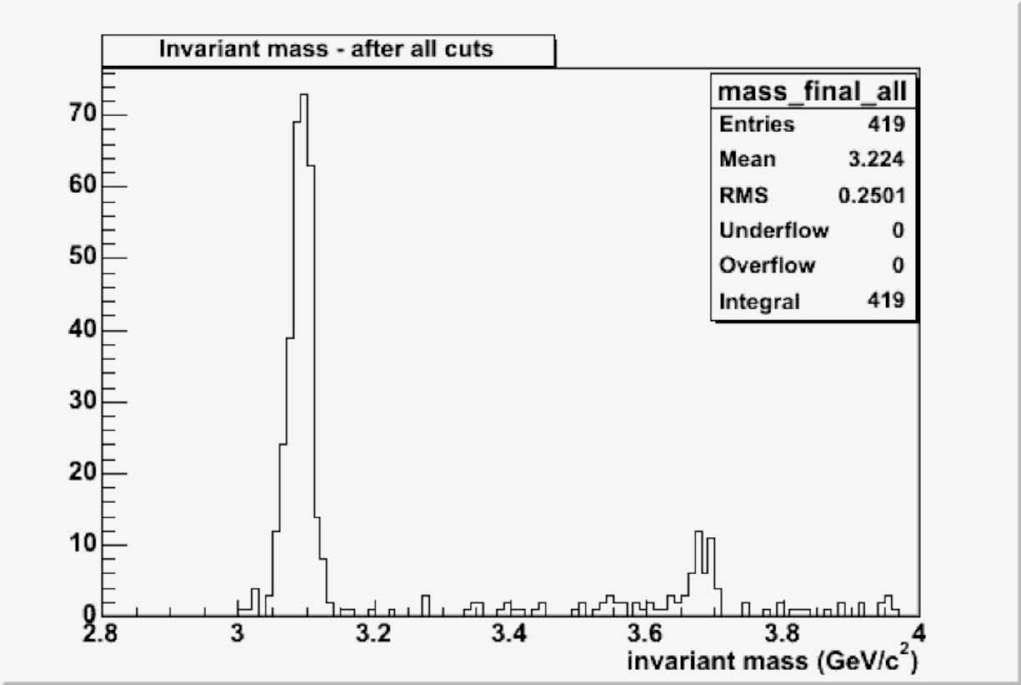}
\includegraphics[height=4.5cm]{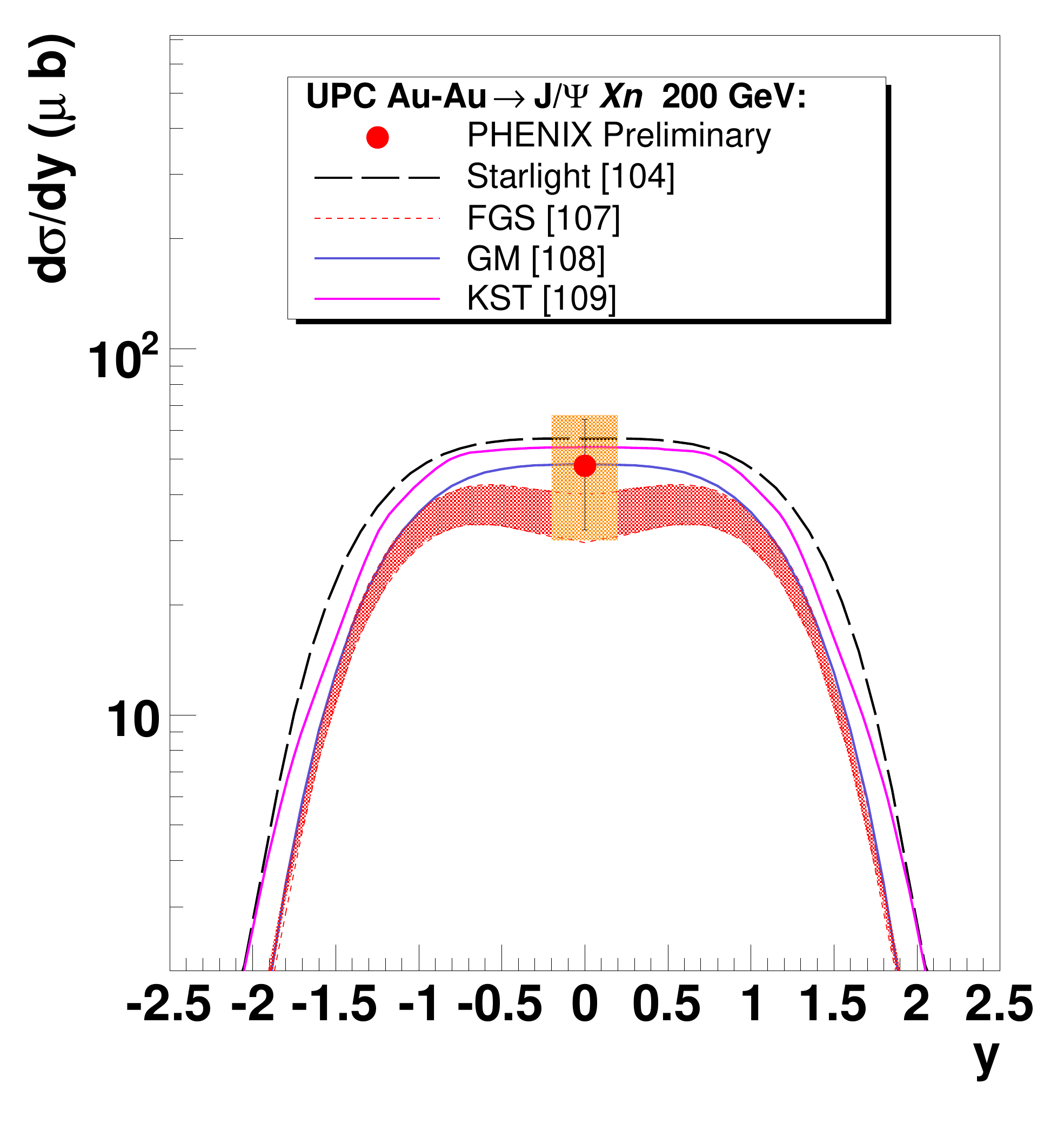}}
\caption{Left: Invariant mass distribution of $\mu^+\mu^-$ pairs measured in exclusive p$\bar{\mbox{p}}$ collisions
by CDF at 1.96 TeV~\protect~\cite{pinfold08}.
Right: Preliminary cross-section of coherent $\jpsi$ production 
at $y$~=~0 in UPC Au-Au at $\sqrtsnn$~=~200~GeV~\cite{dde_qm05} compared 
to various theoretical calculations~\protect~\cite{starlight,strikman05,machado07,kopeliovich07}.}
\label{fig:excl_qqbar}
\end{figure}

The CDF collaboration has recently reported preliminary measurements of exclusive 
$\jpsi$, $\psi'$ and $\ups$ photoproduction in the dimuon decay channel in p-$\bar{\mbox{p}}$ 
collisions at 1.96~TeV~\cite{pinfold08}. The $\ccbar$ states can be very well observed above a small 
dimuon continuum from $\gaga$ interactions (Fig.~\ref{fig:excl_qqbar}, left). The data is compared to photon-pomeron
predictions as implemented in the \str\ Monte Carlo~\cite{starlight} in order to try to pinpoint a 
possible excess which could be indicative of photon-odderon $\jpsi$ production. At higher masses
the $\ups(1S)$ and $\ups(2S)$ are also clearly visible. Similar simulation studies in \pp\ collisions
at the LHC have been carried out by ALICE~\cite{nystrand08} and CMS~\cite{jhollar08}.\\

In Ultra-Peripheral Collisions (UPCs) of heavy-ions the maximum photon energies attainable 
are $\omega_{max}\approx$~3~GeV (100~GeV) at RHIC (LHC). Correspondingly, the 
maximum photon-nucleus c.m. energies are of the order 
$W^{max}_{\gA}\approx$~35~GeV (1~TeV) at RHIC (LHC). Thus, in 
$\gA\rightarrow \jpsi \,(\ups)\, A^{(*)}$ processes, the gluon distribution can be
probed at values as low as $x=M_V^2/W_{\gA}^2\approx 10^{-2} (10^{-4})$.
At low-$x$, gluon saturation effects are expected to reveal themselves through strong 
suppression of hard-exclusive diffraction relative to leading-twist shadowing~\cite{Frankfurt:2005mc}. 
While this suppression may be beyond the kinematics achievable for $J/\psi$ 
photoproduction in UPCs at RHIC, $x\approx 0.01$ and 
$Q^2_{\rm eff} \approx M_V^2/4\approx 3$~GeV$^2$, 
it could be important in UPCs at the LHC~\cite{UPCreport}.\\

The PHENIX experiment has measured $\jpsi$ photoproduction at mid-rapidity 
in Au-Au UPCs at $\sqrtsnn$~=~200~GeV in the dielectron channel~\cite{dde_qm05}. 
Within the (still large) experimental errors, the preliminary $\jpsi$ 
cross-section of $d\sigma/dy|_{|y|<0.5}\,=\,48 \pm 14 \mbox{(stat)}\,\pm 16 \mbox{(syst) }\,\mu b$
is consistent with various theoretical predictions~\cite{starlight,strikman05,machado07,kopeliovich07} 
(Fig.~\ref{fig:excl_qqbar}, right). The band covered by the FGS predictions includes the 
$\jpsi$ cross sections with and without gluon shadowing~\cite{strikman05}. Unfortunately, the current 
statistical uncertainties preclude yet any detailed conclusion regarding the nuclear gluon distribution.

At the LHC energies, the cross section for $\ups(1S)$ photoproduction in UPC \PbPb\ 
at $\sqrtsnn$~=~5.5 TeV is of the order of  150 $\mu$b~\cite{starlight,Frankfurt:2003qy}. 
Inclusion of leading-twist shadowing effects in the nuclear PDFs reduces the yield by 
up to a factor of two, $\sigma_{\ups}$~=~78~$\mu$b~\cite{Frankfurt:2003qy}. 
Even larger reductions are expected in calculations including gluon-saturation 
(Colour Glass Condensate) effects~\cite{Goncalves:2006ed}.
Full simulation studies of input distributions generated with the {\sc starlight} MC~\cite{starlight} 
have shown that ALICE~\cite{nystrand08} and CMS~\cite{cmshitdr} can measure well
$\jpsi\rightarrow \elel$, $\mumu$ and $\ups\rightarrow \elel$, $\mumu$ respectively 
(in different pseudorapidity ranges) in UPCs tagged with neutrons detected in the ZDCs.

\subsection {Bottomonium dissociation at the LHC \protect\small (by David Blaschke)} 

A lot of progresses have been made in the experimental and theoretical 
investigations of charmonium production in heavy-ion collisions, 
see~\cite{Rapp:2008} for a recent review. 
Although CERN-SPS data from the NA50 and NA60 collaborations have provided
a clear evidence for an anomalous $J/\psi$ suppression pattern, an unambiguous 
theoretical approach to the phenomenon is still lacking.
As elements for such an approach are being assembled, it becomes clear that
theoretical models have to be supplemented by measurements of nonperturbative 
phenomena such as to provide baselines for new physics, i.e. for the studies
of heavy quarkonia in hot dense matter as produced in high-energy \AaAa\
collisions.

While the situation with charmonium production becomes more and more complex
and puzzling when comparing SPS and RHIC experiments due to intricate 
intertwining of nonperturbative initial-state, formation-stage and final-state 
effects (see Section~\ref{sec:pA}), the bottomonium spectroscopy accessible in the upcoming LHC 
experiments may offer a much cleaner probe of the physics of dense matter.
This is due to the following specificities of the heavier $b\bar b$ system relative
to the $c\bar c$ system: (i) absence of down-feeding from a heavier quarkonium 
system, (ii) negligible regeneration of bottom quarks from the thermal medium,
(iii) dominance of Coulombic part of the  $b\bar b$ interaction and partonic 
medium, so that plasma screening and thermal dissociation effects shall be 
under better control, (iv) both low-lying states, $\Upsilon (1S)$ and 
$\Upsilon'(2S)$ are bound states in the temperature range $T=1\dots 2 ~T_c$,
and are thus well-separated from the hadronisation stage.

Here we would like to review some first baseline estimates of bottomonium 
spectroscopy for the discussion of upcoming experiments at the LHC and their impact
on the discussion of quark-gluon plasma (QGP) properties.
Due to the above characteristics, bottomonium production at the LHC will allow to
test the kinetic approach to heavy-quarkonium dissociation by thermal activation
in a quark plasma. In this approach the survival probability is given by 
\begin{eqnarray}
S_\Psi &=& S_{\rm HG} \ S_{\rm QGP}  \ S_{\rm nuc}
\nonumber\\
& \simeq & \exp\left(-\int\limits_{T_c}^{T_{fo}}\Gamma_{\rm HG}(T)
\frac{dT}{\dot{T}}\right) \
\exp\left(-\int\limits_{T_c}^{T_{0}}\Gamma_{\rm QGP}(T)
\frac{dT}{\dot{T}}\right) \
\exp\left(-n_N \sigma_{\rm abs} L\right) .
\label{supp-fac}
\end{eqnarray}
where the quarkonium dissociation rate (inverse lifetime) in a QGP is 
determined by the thermally averaged breakup cross sections by (massless)
quark and gluon impact

\begin{equation}
\Gamma_{\rm QGP}(T)=\frac{1}{\tau_{\rm QGP}(T)}=
\sum_{i=q,g}
\frac{1}{2\pi^2}\int_0^\infty \omega^2 d \omega 
\sigma_{(Q\bar Q) i}(\omega) v_{\rm rel} n_i(\omega)~.
\end{equation}

\begin{figure}[!th]
\label{fig:schroedingera}
\centerline{\hspace{1.5cm}\includegraphics[width=0.4\textwidth,angle=-90]{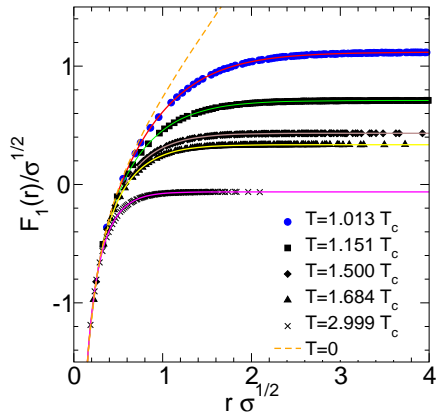}}
\caption{The colour singlet $Q{\bar Q}$ free energy $F(r,T)$ vs.\ $r$ at
different $T$~\cite{Kaczmarek:2003ph,Kaczmarek:2005ui}.}
\end{figure}

\begin{figure}[!th]
\label{fig:schroedingerb}
\hspace{1.5cm}
\includegraphics[width=0.41\textwidth,angle=-90]{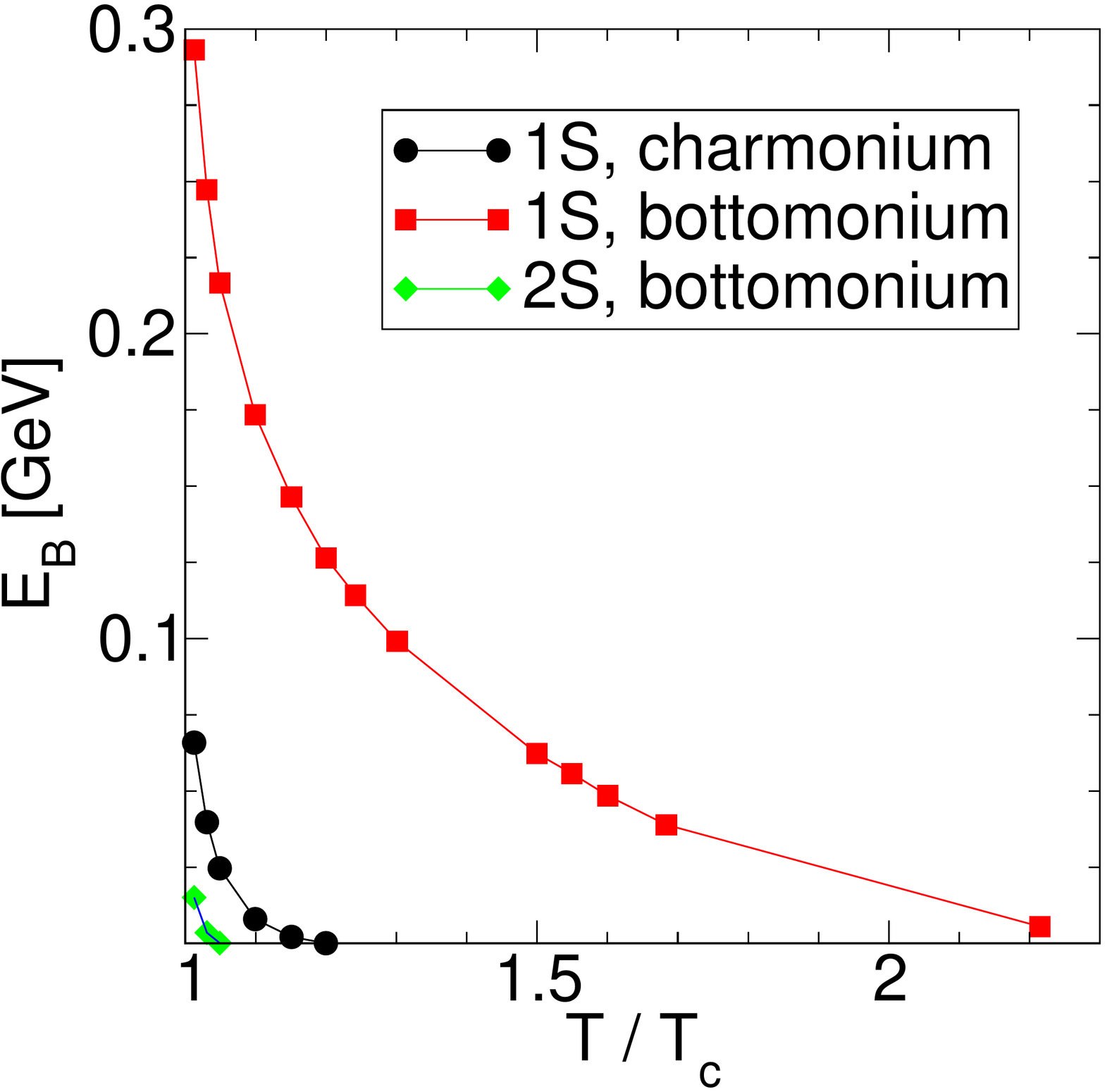}
\hspace{-1.2cm}
\includegraphics[width=0.41\textwidth,angle=-90]{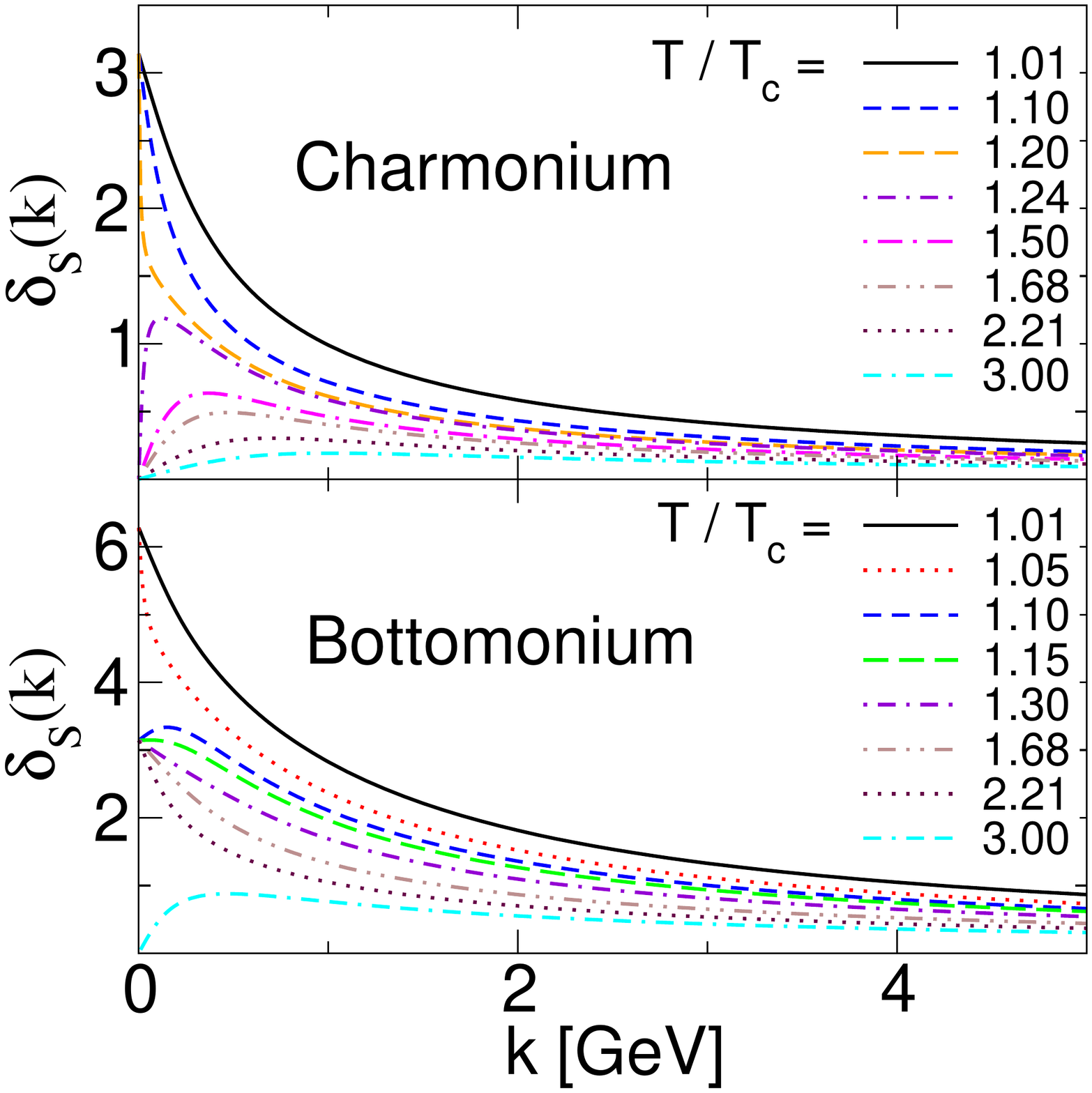}
\caption{
Solutions of the Schr\"odinger equation for heavy
quarkonia with the screened potential identified with the singlet free
energies of the left panel: binding energies (left) and
scattering phase shifts (right), from~\cite{Blaschke:2005jg}. }
\end{figure}

The dominant medium effect on the dissociation rate is given by the temperature
dependence of the binding energy, i.e. the energy necessary to excite the 
quarkonium state to the continuum threshold where it can perform rearrangement 
reactions to open flavor states without energy cost.
These inputs are provided from solutions of the heavy-quarkonium Schr\"odinger
equation with a temperature-dependent heavy-quark potential, see Figs. 
\ref{fig:schroedingera} and \ref{fig:schroedingerb}. 
While the binding energies for $J/\psi$ and $\Upsilon'$ drop to zero for 
temperatures just above $T_c$, the $\Upsilon$ is a bound state up to at least
$2.2~T_c$. 
The absolute value of the binding energy is below the thermal energy of 
the impacting partons $\sim T$ and therefore the $\Upsilon$ production shall 
be dominated by thermal dissociation processes.

\begin{figure}
\hspace{.2cm}
\includegraphics[height=0.4\textwidth,angle=-90]{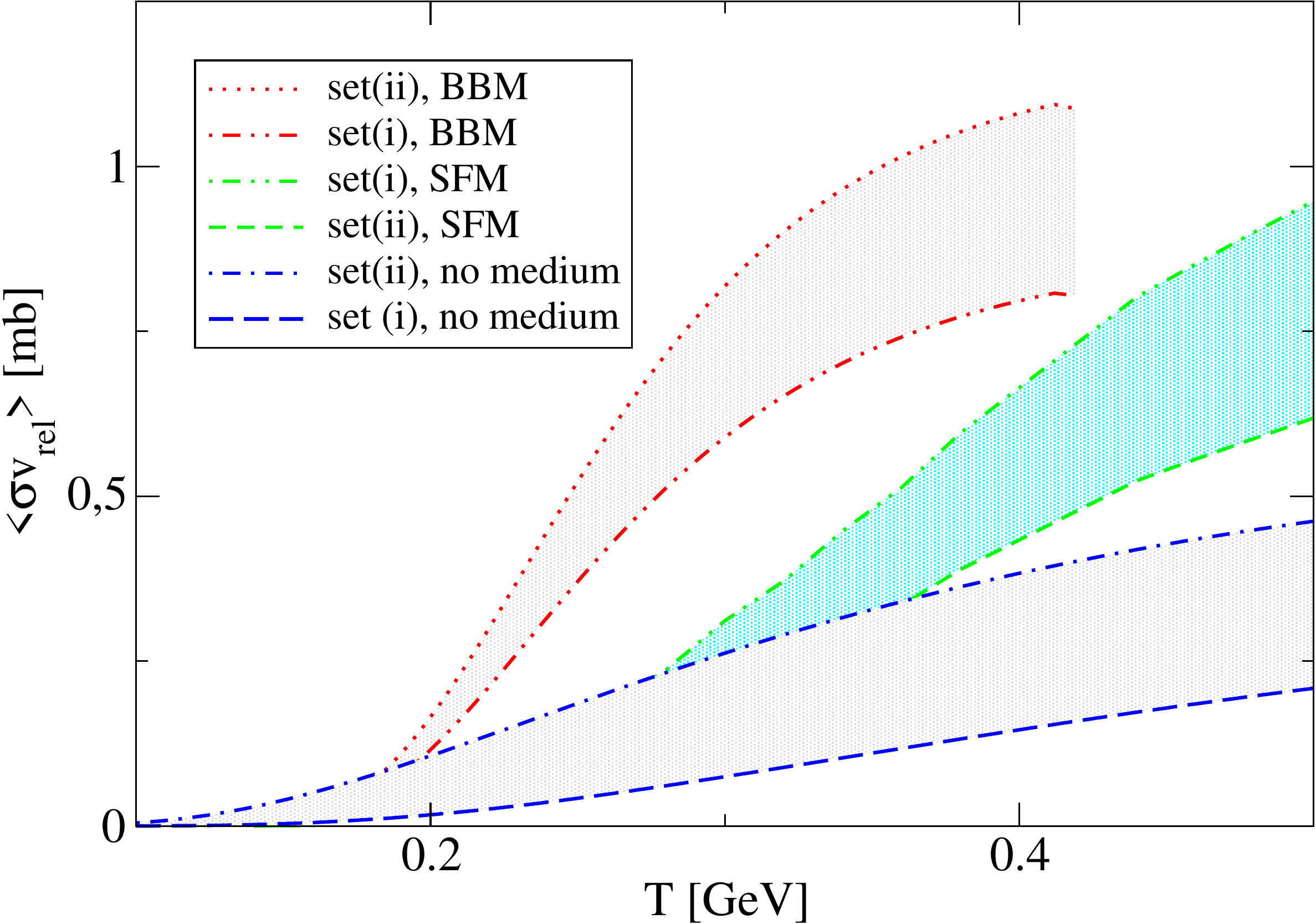}
\includegraphics[height=0.4\textwidth,angle=-90]{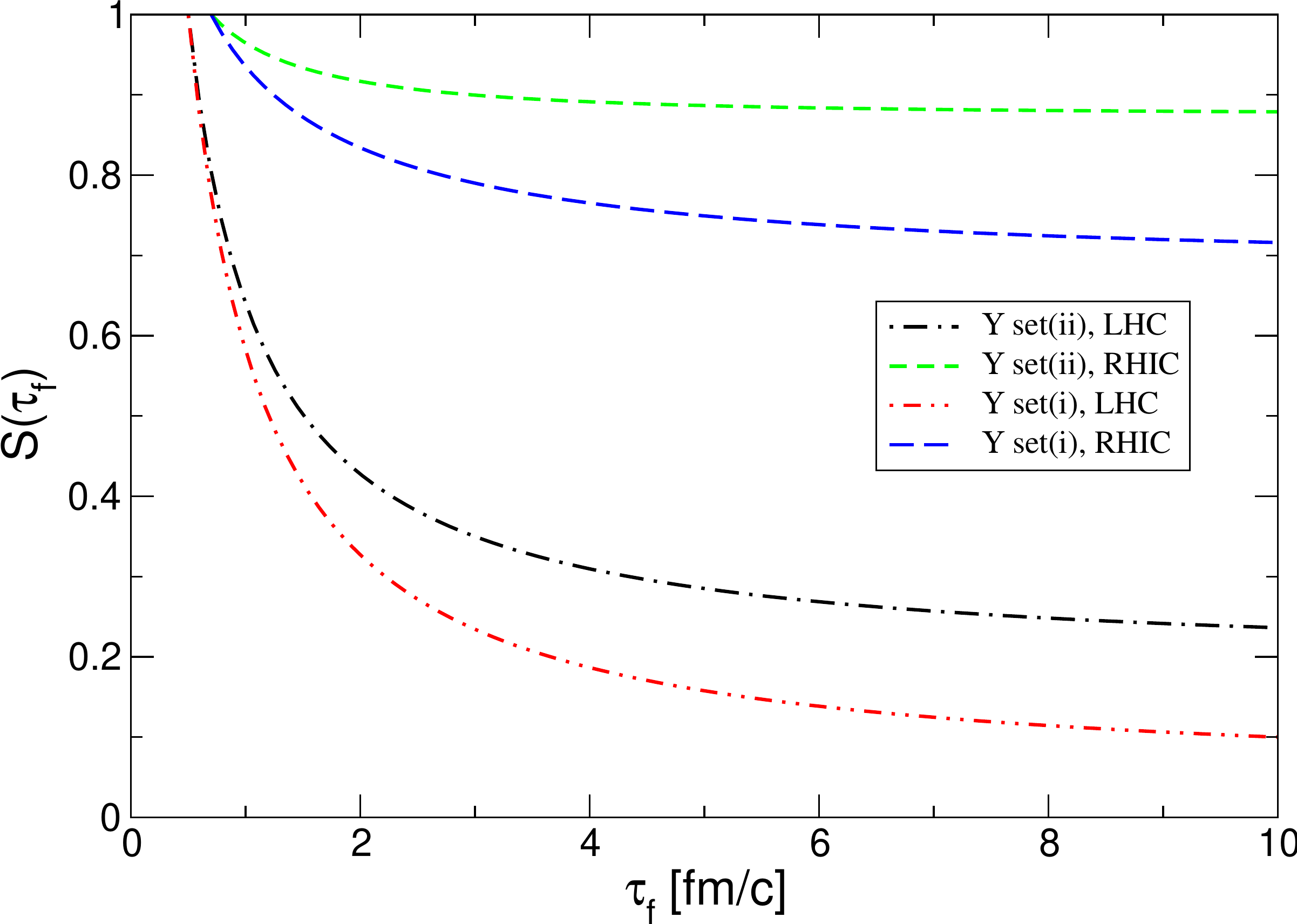}
\caption{Left panel: thermally averaged cross section for $\Upsilon$ 
dissociation by quark impact as a function of the temperature:
BBM and SFM approach give a similar cross section enhancement due to 
the lowering of the breakup threshold, from Ref.~\cite{Blaschke:2004dv}.
Right panel: Survival probability for $\Upsilon$  in a 
longitudinally expanding gluon plasma as a function of the plasma 
lifetime, from Ref.~\cite{Bedjidian:2004gd}. 
\label{fig:dissociation}}
\end{figure}

For the importance of the in-medium modification of the threshold, we want to
refer to processes with quark impact, described by the Bethe-Born model (BBM)
or the string-flip model (SFM), see~\cite{Blaschke:2004dv} and references 
therein, results are shown in Fig. \ref{fig:dissociation}, left panel. 
Corresponding results for gluon-induced processes have recently been discussed
in~\cite{Park:2007zza}.
The right panel of Fig. \ref{fig:dissociation} gives a rough estimate at which
level $\Upsilon$ suppression by thermal dissociation in a QGP is to be 
expected. A more detailed discussion of bottomonium dissociation at RHIC and 
the LHC can be found in~\cite{Grandchamp:2005yw}.

Summarizing, we want to emphasise the role of precise bottomonium spectroscopy 
in \AaAa\ collisions at the LHC for the diagnostics of QGP properties, such as
temperature and lifetime. A key role for the extraction of these properties do 
play microscopic theories for dissociation reactions of the states of the 
$b\bar b$ spectrum with their kinematic dependences.

\section[Bridging the gap between  $pp$ and $AA$ {\rm \protect\small (by Joseph Cugnon)}]{Bridging the gap between $pp$ and $AA$  }
\label{sec:bridging-the-gap}
\vspace*{-.5cm}\hfill  \protect {\bf \small (by~Joseph~Cugnon)}\vspace*{.35cm}\\
This section is devoted to a few remarks on  the actual understanding of quarkonium production 
in heavy-ion collisions in terms of the present knowledge of the properties of quarkonium 
production in free space, of the properties of the quarkonium and on the heavy-ion collision dynamics.

Of course, this survey starts with the status of the theoretical description of the quarkonium 
production in $NN$ collisions. The traditional idea of a gluon fusion  followed by the colour-octet mechanism
 seems to fail to reproduce the CDF data. However, this idea within a colour-singlet picture has been reconciled recently 
with the data for \jpsi~at low $P_T$~\cite{schannelcut} . In a simple language, the interaction between the $c$ and $\bar{c}$ 
quarks has been added. 

As said in Section~\ref{sec:issues-AA}, the conventional belief is that, in heavy-ion 
collisions, quarkonium 
states are produced in the first $NN$ collisions,  similarly as in free space. This is however 
probably not true. At high energy, the incoming heavy-ions should be regarded as fluxes of 
partons. The latter may not be involved  only in the first collisions. The fluxes at the moment 
of interaction may be altered. These ``initial state effects'' on the parton distributions have 
been invoked to describe the $y$ and $P_T$ $J/\psi$ distribution in heavy-ion collisions. It 
seems that these distributions can be described by advocating either the $\ccbar$ 
interaction or the broadening of the $k_T$ distribution of the initial gluons~\cite{KH08}.

Let us now examine the status of the so-called $J/\psi$ suppression. The original idea by Matsui 
and Satz~\cite{Matsui:1986dk} is that the screening of the colour forces inside a plasma leads to the 
eventual dissociation of quarkonium states. Expecting such a plasma in heavy-ion collisions 
at sufficiently large energy, it was predicted that the $J/\psi$ yield should be reduced. 
Since the original yield is not known, the idea is to look at the relative yield as a function 
of the path length in the plasma (in the famous $R_{AA}$ versus $L$ plot). And indeed, 
such a suppression was observed in the NA38 experiment soon after. In the meantime, H\"ufner and
 collaborators showed that such a suppression is obtained in a simple multiple $NN$ collision 
picture, provided an inelastic $J/\psi$-nucleon inelastic cross section of a few mb is 
used~\cite{GE92}. This result shed some trouble for a while, but nobody believes in 
this scenario any more. Evidences for $J/\psi$ suppression has been accumulated, even if they are 
not always consistent. This is not considered as a proof of the existence of the plasma, for 
several reasons. First, the energy spectra of the produced particles do not show temperatures 
above $T_c$ (see the interesting discussion of H. Satz in these proceedings~\cite{satzcontribution}). The alternative 
probes of the plasma do not give clear-cut answers. Furthermore, the $J/\psi$ suppression can be accounted
 for thanks to several alternative scenarios (comovers, meson gas or fluid). 

Interesting developments have taken place in the recent years. There is more and more evidence 
that  quarkonia survive in the plasma up to temperatures of the order of 2$T_c$~\cite{KA}, 
indicating that the transition would not be a pure second order phase transition but rather 
of the Kosterlitz-Thouless type~\cite{MO08}. Therefore, the coupling of a quarkonium state 
and the plasma has to be re-examined. The $J/\psi$-gluon inelastic cross section has been 
re-evaluated recently to take into account this partial screening~\cite{AR05,AR08}. Furthermore, 
the higher-mass resonances are not automatically dissolved. They should increase the $J/\psi$ 
yield. Therefore it has been suggested that the observed $J/\psi$ suppression is largely coming 
from the disappearance of these resonances (the so-called ``direct'' suppression). The evidence 
of that is controversial and there is little hope that experiments at the LHC will clarify the 
issue (see Section~\ref{sec:th-adv-in-pp}).

In the recent years, attention has been put on open~$c$ and open~$b$ production through charmed 
and bottomed mesons and/or jets driven by~$c$ or $b$ quarks. In particular, the $P_T$-dependence 
indicates a strong suppression, contrasting with the small suppression of quarkonium (even no 
suppression at large $P_T$; incidentally, the interest in the $P_T$-dependence of $J/\psi$ 
suppression has been so luckily revived, this variable being less ambiguous than the $L$ variable 
or equivalent). This has raised a strong interest in the energy loss of a heavy quark in the 
plasma. According to some authors~\cite{DO01,MI,KH08}, the radiative energy loss in colour fields 
may lead to an upper bound of the final parton energy, expressed as

\begin{equation}
E_{bound}= \frac{2 \pi}{\sqrt{\lambda}} \frac{m^4}{F^2} \frac{1}{L}
\end{equation}
 where $m$ is the mass of the quark, $L$ is the path length and $F$ is the colour force. This 
results from the so-called AdS/CFT correspondence and crucially depends upon the  fact that 
the specific energy loss is proportional to the square of the energy in such an approach. 
From reasonable estimates, the previous equation leads to 

\begin{equation}
E_{bound} ({\rm GeV}) \leq  \frac{ \xi}{L ({\rm fm})}
\end{equation}
 where $ \xi$=1 for $c$ and 14 for $b$ quarks. If this is true, heavy-quark production will be 
limited to surface emission only.  In addition, recent measurements indicate that $c$ and 
$b$ quarks participate to the flow, which even complicates the picture of the propagation 
inside the plasma. Definitely, this issue warrants further investigation.

In conclusion, if there has been undeniable theoretical progress on various aspects of the 
production of quarkonium in the recent years, one has to admit that our  understanding is 
still far from complete.

\section{Conclusion}

On the verge of the start-up of the LHC, an overview of the current
knowledge of heavy-quarkonium production reveals a situation that is 
somehow still not satisfactory. In proton-proton collisions, the interest 
of this process stems from the simple observation that the rather large scale 
introduced by the heavy-quark mass allows a separation between perturbative
and non-perturbative physics, opening up the way to a first principle
description.  However, given the present data, the effectiveness of 
such an approach is yet to be confirmed.

Concerning nucleus-nucleus collisions, the status of $J/\psi$
suppression as an indicator of the formation of the quark-gluon plasma
(QGP) is also becoming more shaky. There has been a great effort to
use proton-nucleus collisions to evaluate the so-called Cold Nuclear
Matter (CNM) effects on the $J/\psi$ suppression in the recent years.

Yet the situation appeared more complicated than expected concerning
the nuclear parton distribution functions (nPDFs) at low $x$ and their
dynamical evolution. Although novel experimental observables, like e.g.
quarkonium photoproduction in ultra-peripheral nucleus-nucleus collisions,
provide a new handle to shed light on these issues.
The influence of the QGP on the suppression is
not really transparent: the understanding of the screening of colour
charges above the critical temperature, of the coupling of the heavy
quarkonium to the plasma and of its time evolution seems to be less
solid than a few years ago.

On the other hand, several aspects are promising, most of them being
related to the advent of the LHC beams. As for heavy-quarkonium
production in proton-proton collisions, the possibility of bottomonium
production, of measurements of polarisation and associated production
of $J/\psi$ and of $\Upsilon$ with a heavy-quark pair will severely
constrain the current theoretical models. As for proton-nucleus
collisions the analysis of the \dAu\ data at RHIC and the set-up of an
elaborated \pA\ program are certainly of importance and promising of
future programs.

Finally, for nucleus-nucleus collisions, the LHC will offer new
possibilities: higher temperatures of the plasma and production of
$\Upsilon$ states, which are expected to clarify the
analysis. Theoretical advances concerning the interaction of the heavy
quarkonia with the hot plasma, the slowing down of heavy quarks and
the space-time evolution of the plasma are expected. Those advances
are anyhow necessary.

\section*{Acknowledgments}

We thank J.R. Cudell and S. Peign\'e for their participation to the round table. We would like
also to thank M. Calderon, F. Fleuret, R. Granier de Cassagnac, B. Kopeliovich, U. Langenegger,
G. Martinez, A. Meyer, H.J. Pirner and K. Schweda for useful and motivating discussions.


\end{document}